%% file: paper.tex
\documentclass[fleqn,usenatbib]{mnras}

\usepackage[T1]{fontenc}
\usepackage{ae,aecompl}
\usepackage{multirow}
\usepackage{dirtytalk}

\usepackage[normalem]{ulem}

\usepackage{graphicx}	
\usepackage{amsmath}	
\usepackage{amssymb}	
\usepackage{pdfpages}   
\usepackage{hyperref}
\hypersetup{
pdfauthor={S.~Hackstein,
           M.~Brueggen,
           F.~Vazza,
           L.~F.~S.~Rodrigues (orcid.org/0000-0002-3860-0525)
           },
    pdftitle={PrEFRBLE}
}

\graphicspath{{figures/}}

\newcommand{\dd}{\text{d}}
\newcommand{\bfrac}[2]{\left( \frac{#1}{#2} \right)}
\newcommand{\ufrac}[2]{\bfrac{#1}{\rm #2}}
\newcommand{\avg}[1]{\langle #1 \rangle}

\newcommand{\PreFRBLE}{{\sc PrEFRBLE}}
\newcommand{\PreFRBLElong}{Probability Estimates for Fast Radio Bursts to model Likelihood Estimates}
\newcommand{\primordial}{{\it primordial}}
\newcommand{\astrophysical}{{\it astrophysical}}

\newcommand{\Rodrigues}{{\it Rodrigues19}}

\newcommand{\unitDM}{{\rm\ pc \ cm^{-3}}}
\newcommand{\unitRM}{{\rm\ rad \ m^{-2}}}
\newcommand{\unitSM}{{\rm\ kpc \ m^{-20/3}}}

\newcommand{\RMEG}{\text{RM}_{\rm EG}}

\newcommand{\RMIGM}{\text{RM}_{\rm IGM}}
\newcommand{\RMInter}{\text{RM}_{\rm Inter}}

\newcommand{\RMHost}{\text{RM}_{\rm host}}

\newcommand{\RM}{\text{RM}}
\newcommand{\DMobs}{\text{DM}_{\rm obs}}
\newcommand{\DMEG}{\text{DM}_{\rm EG}}
\newcommand{\DMMW}{\text{DM}_{\rm MW}}
\newcommand{\DMIGM}{\text{DM}_{\rm IGM}}

\newcommand{\DMHost}{\text{DM}_{\rm host}}

\newcommand{\DM}{\text{DM}}

\newcommand{\SMIGM}{\text{SM}_{\rm IGM}}
\newcommand{\SMInter}{\text{SM}_{\rm Inter}}

\newcommand{\SMeff}{\text{SM}_{\rm eff}}

\newcommand{\SM}{\text{SM}}

\newcommand{\tauIGM}{\tau_{\rm IGM}}
\newcommand{\tauInter}{\tau_{\rm Inter}}

\newcommand{\fIGM}{f_{\rm IGM}}
\newcommand{\zFRB}{z_{\rm FRB}}
\newcommand{\zInter}{z_{\rm Inter}}
\newcommand{\NInter}{N_{\rm Inter}}
\newcommand{\piInter}{ \pi_{\rm Inter} }
\newcommand{\Deff}{D_{\rm eff}}
\newcommand{\rdiff}{r_{\rm diff}}
\newcommand{\Bayes}{\mathcal{B}}
\newcommand{\Model}{\mathcal{M}}

\newcommand{\ana}{A\&A}

\newcommand{\galform}{\textsc{Galform}}
\newcommand{\magnetizer}{\textsc{Magnetizer}}

\title[PrEFRBLE]{Redshift estimates for fast radio bursts and implications on intergalactic magnetic fields}

\author[Hackstein et al.]{
{S. Hackstein$^{1}$}\thanks{E-mail: stefan.hackstein@hs.uni-hamburg.de},
{M. Br\"uggen$^{1}$},
{F. Vazza$^{1,2}$},
{L. F. S. Rodrigues$^{3}$}
\\
$^{1}$Hamburger Sternwarte, University of Hamburg, Gojenbergsweg 112, 21029, Germany\\
$^{2}$University of Bologna,  Department of Physics and  Astronomy,
Via Gobetti 93/2, I-40129, Bologna, Italy;\\$^{\phantom{22}}$Istituto di Radioastronomia, INAF, Via Gobetti 101,40129 Bologna, Italy\\
$^{3}$Department of Astrophysics/IMAPP, Radboud University, Postbus 9010,
6500 GL, Nijmegen, Netherlands
}

\pubyear{2020}

\begin{document}
\label{firstpage}
\pagerange{\pageref{firstpage}--\pageref{lastpage}}
\maketitle

\begin{abstract}

\textit{Context}: Fast Radio Bursts are transient radio pulses from presumably compact stellar sources of extragalactic origin.
With new telescopes detecting multiple events per day, statistical methods are required in order to interpret observations and make inferences regarding astrophysical and cosmological questions.

\textit{Purpose:} We present a method that uses probability estimates of fast radio burst observables to obtain likelihood estimates for the underlying models.

\textit{Method}: Considering models for all regions along the line-of-sight, including intervening galaxies, we perform Monte-Carlo simulations to estimate the distribution of the dispersion measure, rotation measure and temporal broadening. Using Bayesian statistics, we compare these predictions to observations of Fast Radio Bursts.

\textit{Results}: By applying Bayes theorem, we obtain lower limits on the redshift of Fast Radio Bursts with extragalactic $\DM \gtrsim 400 \unitDM$.
We find that intervening galaxies cannot account for all highly scattered Fast Radio Bursts in FRBcat, thus requiring a denser and more turbulent environment than a SGR 1935+2154-like magnetar.
We show that a sample of $\gtrsim 10^3$ unlocalized Fast Radio Bursts with associated extragalactic $\RM\geq 1 \unitRM$ can improve current upper limits on the strength of intergalactic magnetic fields.
\end{abstract}





\input{introduction}

\input{measures}

\input{prefrble}

\input{models}

\input{applications}

\input{discussion}


\section*{Acknowledgements}
This work was funded by the Deutsche Forschungsgemeinschaft (DFG) under grant BR2026/25.
LFSR acknowledges funding from the European Research Council (ERC) under the European Union's Horizon 2020 research and innovation programme (grant agreement No 772663).
SH thanks David Gardenier for helpful remarks and for making publicly available the software package {\sc FRBpoppy}.
Furthermore, SH thanks Rainer Beck, Laura Spitler, Sui Ann Mao, Maja Kierdorf, Charles Walker, Hsiu-Hsien Lin and Pranjal Trivedi for interesting and fruitful discussions on many aspects of this work. LFSR thanks Luke Chamandy for useful discussions.
The whole group thanks the referee, Maxim Pshirkov, for interesting remarks.

Our cosmological simulations were performed with the {\sc ENZO} code (http://enzo-project.org), under project HHH38 and HHH42 at the J\"ulich Supercomputing Centre (P.I. F. Vazza). 
FV acknowledges financial support from the ERC  Starting Grant \say{MAGCOW}, no. 714196.
We thank Jenny Source and Stefan Gottl\"ober for providing us with the initial conditions and for their help in implementation. 

We also acknowledge the use of computational resources at the Rechenzentrum of the University of Hamburg

\section*{Data availability}

The data underlying this article are available in the PrEFRBLE repository, at \href{http://doi.org/10.5281/zenodo.3862636}{http://doi.org/10.5281/zenodo.3862636}




\bibliographystyle{mnras}
\bibliography{cites} 

\begin{table}

\begin{tabular}{l|c|l}
Symbol & units & description \\
\multirow{2}{*}{ $L(v|\Model)$ } & \multirow{2}{*}{-} & model likelihood of measure $v$ \\ & & ~in  case of model $\Model$ \\
$\pi(\Model)$ & - & prior of model $\Model$\\
\multirow{2}{*}{ $P(\Model|v)$ } & \multirow{2}{*}{-} & posterior of model $\Model$ \\ & & ~in face of observed measure $v$ \\ 
$\Bayes(v|\Model_1, \Model_2)$ & - & Bayes factor, Eq. \eqref{eq:Bayes} \\
RM & $\rm rad~m^{-2}$ & Faraday rotation measure \\
DM & $\rm pc~cm^{-3}$ & dispersion measure \\
SM & $\rm kpc~m^{-20/3}$ & scattering measure \\
$\tau$ & $\rm ms $ & temporal smearing \\
$B$ & $\rm  \mu G$ & magnetic field strength \\
\multirow{2}{*}{$B_0$} & \multirow{2}{*}{$\rm  nG$} & comoving primordial \\ & & ~magnetic field strength \\
$\rho$ & $\rm g~cm^{-3}$ & baryonic gas density  \\
$\alpha$ & - & exponent of $|B|\propto\rho^\alpha$ relation \\
$M_\star$ & $M_\odot$ & total stellar mass of galaxy \\
$\Phi$ & $\rm Mpc^{-3} $ & galaxy stellar mass function  \\
$\epsilon$ & - & properties of galaxy \\
$z$ & - & redshift \\
$n_{\rm gal}$ & $\rm Mpc^{-3}$ & number density of galaxies \\
$r_{\rm gal}$ & $\rm kpc$ & disk size of galaxy model \\
$r_{1/2}$ & $\rm kpc$ & galaxy half mass radius \\
$d_H$ & $\rm Gpc$ & hubble radius \\
$C_N^2$ & $\rm m^{-20/3}$ & turbulence per unit length  \\
$L_0$ & $\rm pc$ & outer scale of turbulence \\
$l_0$ & $\rm pc$ & inner scale of turbulence \\
$n_e$ & $\rm cm^{-3}$ & electron number density \\
$\nu$ & $\rm Hz$ & frequency  \\
$\lambda$ & $\rm cm$ & wavelength \\
$D$ & $\rm Gpc$ & angular diameter distance \\
$\Deff$ & $\rm Gpc$ & effective lense distance 
\end{tabular}
\caption{List of symbols used in the paper.}
\label{tab:symbols}
\end{table}

\begin{table}
\begin{tabular}{l|l}
Subscript & description \\

IGM & intergalactic medium \\
FRB & source of FRB \\
Host & host galaxy \\
Inter & intervening galaxy \\
Local & local source environment \\
MW & Milky Way \\
gal & galactic \\
obs & observed \\
res & residual \\
eff & effectively observed \\
$L$ & lensing medium
\end{tabular}
\caption{List of subscripts used in the paper.}
\label{tab:subscripts}
\end{table}

\input{appendix}



\label{lastpage}
\end{document}

%% file: introduction.tex
\section{Introduction}

\subparagraph{Fast Radio Bursts}

Fast radio bursts (FRBs) are millisecond transient sources at $\approx 1\rm~GHz$ with very high luminosities, first discovered by \citet{lorimer2007}.
Their observed dispersion measure (DM) often exceeds the contribution of the Milky Way (MW), suggesting an extragalactic origin.
FRBs have the potential to help answer many long-lasting astrophysical and cosmological questions \citep[e.g.][for reviews]{katz2016,ravi2019fast,Petroff2019review}, provided theoretical predictions can be tested against observations.
For this purpose, we present a Bayesian framework to constrain models of FRB sources as well as the different regions along their lines-of-sight (LoS): the intergalactic medium (IGM), the host and intevening galaxies as well as the local environment of the progenitor.

\subparagraph{FRB progenitor}

Numerous models have been put forward that explain the origin of FRBs. These models are collected in the living theory catalog\footnote{\href{https://frbtheorycat.org/}{frbtheorycat.org}} \citep{platts2018livingcat}.
Many models assume that flares of young neutron stars cause shock waves in the surrounding medium, where gyrating particles emit coherent emission \citep{Popov2010,lyubarsky2014,murase2016,beloborodov2017,metzger2019}.
Cataclysmic events usually consider interactions of magnetic fields during the merger of two compact objects, e. g. two neutron stars \citep{wang2016}, or during the collapse of a single object, e. g. neutron star to black hole \citep{fuller2015}.
The search for an FRB counter-part proves elusive \citep{scholz2016,bhandari2017,scholz2017,xi2017}, except for the possible detection of a transient $\gamma$-ray counterpart to FRB131104 \citep{delaunay2016} and a $\gamma$-ray burst with spatial coincidence to FRB171209 \citep{wang2020grb}, which both point to magnetars \citep[see also][]{metzger2017,zanazzi2020,li2020comparative}.
Furthermore, the recent detection of a X-ray flare from Galactic magnetar SGR 1935+2154, accompanied by a radio burst of millisecond duration consistent with cosmological FRBs \citep{collaboration2020bright,bochenek2020fast,lyutikov2020fast,mereghetti2020integral}, provides strong evidence for magnetars as sources of FRBs, though these are required to be different from Galactic magnetars \citep{margalit2020implications}.

\subparagraph{FRBs as cosmological probes}

The use of FRBs as cosmological probes has been discussed in several papers.
FRBs might be used to constrain the photon mass \citep{wu2016}, violations of Einstein's equivalence principle \citep{wei2015,tingay2016}, Dark Matter \citep{munoz2016, sammons2020constraints, liao2020constraints} and cosmic curvature \citep{li2018}.
Several publications discuss the use of DM-redshift relation of either FRBs associated with $\gamma$-ray bursts or localized FRBs to constrain the equation of state of dark energy as well as other cosmological parameters \citep{zhou2014,gao2014,yang2016,walters2018,wu2020new}.
\citet{wucknitz2020cosmology} show how to use gravitationally lensed repeating FRBs to constrain cosmological parameters (see also \citet{wei2018, jaroszynski2019}).

\subparagraph{FRB localization}

Most methods to use FRBs as cosmological probes requires the localisation of a large number of FRBs.
However, the localization of sources of short-duration signals without known redshift is difficult \citep{eftekhari2017,mahony2018,marcote2019,prochaska2019uncovering}.
The current sample of known host galaxies of five localized FRBs includes massive as well as dwarf galaxies, with some showing high, others low rates of star formation \citep{tendulkar2017host,ravi2019fast,bannister2019single,prochaska2019low,marcote2020}.

Here, we show how to use unlocalized FRBs with reasonable assumptions on their intrinsic redshift distribution to test models of FRBs and the intervening matter.

\subparagraph{FRB redshift distribution}

Several researchers have tried to infer the intrinsic redshift distribution of FRBs either by modelling the distribution of DM and other FRB properties with analytical or Monte-Carlo methods \citep{bera2016,caleb2016,gardenier2019synthesizing}, or by performing a luminosity-volume test  \citep{locatelli2018}.
They conclude that data sets from different telescopes disagree on the redshift distribution.

There has been previous work to estimate the redshift of individual FRB sources based on their DM \citep{dolag2015,Niino2018,luo2018,Walker2018,Pol2019}.
The observed DM is dominated by the long scales of the IGM already at low redshift, $z\lesssim 0.1$.
However, possible contributions by high-density regions (e.g. halos of galaxies) or the local environment of the source can bias the use of DM to infer the redshift of the source $\zFRB$.
Thus earlier work has concluded that only upper limits on $\zFRB$ can be derived based on DM.

\subparagraph{FRBs as probe for intergalactic magnetic fields}

Some FRBs show high levels of linear polarization, up to 100 per cent \citep{michilli2018,day2020high}.
Their associated Faraday rotation measure (RM) contains information on the traversed magnetic field.
\citet{akahori2016} and \citet{vazza2018frb} show that DM and RM of FRBs potentially signal information about the intergalactic magnetic field (IGMF).
However, so far a detailed investigation of the combined contribution of all other regions along the line of sight is missing.

Magnetic fields in galaxies \citep[e. g.][]{beck2016} have been investigated mainly using synchrotron emission via Faraday rotation of background radio sources or RM Synthesis. 
However, due to limited sensitivity and angular resolution, observing galaxies and their properties becomes increasingly difficult at high redshifts \citep{bernet2008,mao2017}.

Magnetic fields in clusters are of the order of $0.1 - 10 ~\rm \mu G$ \citep[e. g. ][]{vanWeeren2019}.
However, the strength and shape of IGMFs in the low-density Universe is still poorly constrained \citep[e. g.][]{taylor2011,dzhatdoev2018}.
Current limits range from $B\lesssim 4.4 \times 10^{-9} \rm ~G$ comoving \citep{Planck2015PMF} to $B \gtrsim 3\times 10^{-16} ~\rm G$ \citep{neronov2010}.

In \citet{hackstein2019}, we developed a framework to investigate the combined contribution to RM from all regions along the LoS.
We could show that this allows us to tell apart extreme models for the origin of IGMFs.
Here, we refine the modelling of IGMFs and investigate how many unlocalized FRBs observed with RM are required  to improve current constraints on IGMFs.

\subparagraph{Intervening galaxies}

The LoS to a source at cosmological distances has significant chances to traverse an additional galaxy between host galaxy and the MW \cite[e. g. ][]{macquart2013temporal}.
Due to the lower redshift, contributions to the RM are probably even higher than for the host galaxy, limiting our ability to probe IGMFs.
However, intervening galaxies are expected to dominate temporal smearing $\tau$ due to the ideal position of the high-density plasma lense \citep{macquart2013temporal}.
Here we investigate the use of $\tau$ to identify LoS with intervening galaxies.

For this purpose, we have created the open-source python software package \PreFRBLE \citep{PreFRBLEzenodo}, using a framework of Bayesian inference, similar to \citet{Luo_2020} and \citet{Macquart_2020}.
The observational measures investigated in this paper are shortly discussed in Sec.~\ref{sec:measures}.
In Sec.~\ref{sec:prefrble}, we summarize the statistical methods used in \PreFRBLE.
The different models are explained in Sec.~\ref{sec:models}.
A few possible applications of \PreFRBLE ~using FRBs in FRBcat are presented in Sec. \ref{sec:applications}:
Identification of intervening galaxies is discussed in Sec.~\ref{sec:identify}.
We estimate the host redshift of unlocalized FRBs in Sec.~\ref{sec:redshift}.
In Sec.~\ref{sec:IGMF} we show how to infer the IGMF from DM and RM of unlocalized FRBs.
Finally, we conclude in Sec.~\ref{sec:discussion}.
A list of all symbols used throughout the paper is shown in Tab. \ref{tab:symbols}.
Explanations of subscripts can be found in Tab. \ref{tab:subscripts}.

%% file: measures.tex

\section{Observables}
\label{sec:measures}


\subsection{Dispersion measure}
\label{sec:DM}

When propagating through plasma, radio waves are dispersed, causing a delay in arrival time that scales with the squared wavelength \citep[e. g.][]{mcquinn2013}.
This delay is quantified by the frequency-independent DM, defined as the free electron column density
\begin{equation}
\text{DM} = \int_0^d \ufrac{n_e}{cm^{-3}} ~\ufrac{\dd l}{pc} \unitDM ,
\label{eq:DM}
\end{equation}
i. e. the number of free electrons per unit volume $n_e$ along the LoS to distance $d$.
Due to their large volume filling factor in the cosmic web, filaments, walls and voids contribute most of the DM by the IGM, while galaxy clusters account for only $\sim 20$ per cent of $\DMIGM$  \citep{zhu2018scattering}.
Thus, DM can be used to infer the distance to the FRB \citep{dolag2015,Niino2018,luo2018,Walker2018,Pol2019}.


\subsection{Faraday rotation measure}
\label{sec:RM}
Linearly polarised radio waves that travel through a magnetised plasma experience a rotation in their polarization angle.  
This is quantified by the frequency-independent RM, defined as the column density of free electrons times magnetic field along the LoS $B_\parallel$, 
\begin{equation}
\text{RM} \approx 0.81 \int_d^0 \ufrac{n_e}{cm^{-3}} ~\ufrac{B_\parallel}{\mu G} ~\ufrac{\dd l}{pc} \unitRM .
\label{eq:RM}
\end{equation}
However, significant contributions to the RM are expected from all regions along the LoS \citep[e. g.][]{hackstein2019}, which complicates their interpretation.

RM can be positive and negative, thus contributions from separate regions may cancel each other out.
This is considered in the numerical computation of results for the full LoS (Eq. \eqref{eq:convolve}).


\subsection{Temporal smearing}
\label{sec:tau}
Inhomogeneities in a turbulent plasma can partly scatter radio waves off and back onto the LoS.
Multipath propagation creates a partial delay of the signal, causing temporal smearing $\tau$ of short pulses, as well as angular scattering $\theta_{\rm scat}$ of the observed signal.
However, $\tau$ strongly depends on the wavelength of the scattered wave. 
It can be calculated by the frequency-independent scattering measure (SM),
which is defined as the path integral over the amplitude of the turbulence per unit length, $C_N^2$, \citep{macquart2013temporal}
\begin{equation}
\SM = \int_l^{l+\Delta l} \ufrac{C_N^2}{1~m^{-20/3}} \ufrac{\dd l}{1~kpc} ~\rm kpc~ m^{-20/3} .
\label{eq:SM}
\end{equation}
For objects that are part of the Hubble flow, it is convenient to define the effective scattering measure 
\begin{equation}
\SMeff = \int\limits_l^{l+\Delta l} C_N^2 (1+z)^{-2} \dd l ,
\label{eq:SMeff}
\end{equation}
that refers all quantities back to the observers frame.
\citet{macquart2013temporal} give an estimate for the amplitude of Kolmogorov turbulence inside of galaxies
\begin{equation}
C^2_{N,gal} = 1.8 \times 10^{-3} \bfrac{n_e}{10^{-2} \rm~cm^{-3}}^2 \bfrac{L_0}{0.001 \rm~pc}^{-2/3} \rm~m^{-20/3} .
\label{eq:c_ngal}
\end{equation}
We follow the argument of \citet{macquart2013temporal} and assume a fully modulated electron density, $\delta n_e \approx n_e$, and that the power spectrum of turbulence follows a power law with index $\beta$, hence $C_N^2 \propto \avg{\delta n_e^2} L_0^{\beta - 3} = \avg{n_e}^2 L_0^{\beta - 3}$.
For a power law with sufficient range, i. e. inner scale $l_0 \ll L_0$, SM is determined by the outer scale of turbulence $L_0$.
\\

Future observations of FRBs may provide observed SM by comparing $\theta_{\rm scat}$ and $\tau$ at different frequencies.
However, FRBs available in the FRBcat catalog \citep{FRBCAT} provide only observed $\tau$ for the dominant frequency of the burst.
Extracting SM from $\tau$ requires assumptions on the redshift of source and scattering medium.
Hence, by directly predicting $\tau$ instead of SM, comparison to observations relies on fewer assumptions.
\\

According to \citet{macquart2004}, the temporal smearing can be approximated by a thin screen approximation, even for media extended along the LoS.
For radio signals with wavelength $\lambda_0$, scattered by a medium at redshift $z_L$, \citet{macquart2013temporal} provide a numerical expression for the scattering time
\begin{equation}
\tau = 1.8 \times 10^8 {~\rm ms} \bfrac{\lambda_0}{1 \rm~m}^{\frac{22}{5}} (1+z_L)^{-1} \bfrac{\Deff}{1\rm~Gpc} \bfrac{\SMeff}{\unitSM}^{\frac{6}{5}} 
\label{eq:tau}
\end{equation}
with effective lensing distance $\Deff = \frac{D_L D_{LS}}{D_S}$, i. e. the ratio of angular diameter distances observer to source $D_S$, observer to scattering medium $D_L$ and medium to source $D_{LS} \neq D_S - D_L$.
This result requires that $l_0$ is smaller than the length scale of plasma phase fluctuations $r_{\rm diff}$.
Numerical tests show, that for the frequencies of FRBs considered in this paper, $r_{\rm diff} > l_0\approx 1 \rm~AU$ always (See App. \ref{sec:rdiff}).

We compute results for $\lambda_0 = 0.23 \rm~m$, corresponding to a frequency $\nu \approx 1300 \rm~MHz$.
Temporal scattering at other wavelengths, $\lambda$, can simply be computed in post-processing, by applying a global factor of $(\lambda / \lambda_0)^{\frac{22}{5}}$. 
Considering that $\SMeff \propto (1+z)^{-2}$, Eq. \eqref{eq:tau} implies temporal scattering occurring within the host galaxy, computed once assuming $\zFRB=0$, e. g. for the redshift independent model of the local environment, scales with source redshift as $\tau(\zFRB) \propto (1+\zFRB)^{-\frac{17}{5}}$.

%% file: prefrble.tex

\section{\PreFRBLE}
\label{sec:prefrble}

\PreFRBLE\footnote{\href{https://github.com/FRBs/PreFRBLE}{github.com/FRBs/PreFRBLE}}, \say{\PreFRBLElong}, is an open-source Python software package designed to quantify predictions for the RM, DM and SM of FRBs and compare them to observations \citep{PreFRBLEzenodo}.
The results can be used to obtain estimates of the likelihood of models of progenitors of FRBs as well as the different regions along their LoS.

\subparagraph{Model likelihood}

We model the contribution of individual models using Monte-Carlo simulations.
The distribution of predicted measures $v(\theta)$, sampled randomly according to a
prior distribution $\pi(\theta)$ of model parameters $\theta$,
reflects the expected likelihood to observe a given measure, $L(v|\Model)$, to which we refer as model likelihood.
$L(v|\Model)$ is also known as model evidence or marginal likelihood function, as it is marginalised
over any model parameters, i.e. 
\begin{equation}
L(v|\Model)=\int L'(v|\Model,\theta)\pi(\theta) ~\dd \theta.
\end{equation}
A detailed description of the Monte-Carlo simulations for the individual models are presented in Sec. \ref{sec:models}.

We often use a logarithmic range for the values described by $L$, resulting in an uneven binning of results.
When visualising the model likelihood, we either plot the
complementary cumulative likelihood, $L(>x) = \int_x^{\infty} L(x) ~\dd x$, or
the product, $L(x)\cdot x$, which is a physical value and not affected by binning.

\subparagraph{Combine models of separate regions}

We consider the contribution from the following regions along the LoS of FRBs: the local environment of the progenitor, the host and intervening galaxies, the IGM.
However, we neglect the foregrounds of the MW and the Earth's ionosphere. All of these regions are described by separate models.
When provided in the form of likelihoods $L(\text{measure}|\text{model})$ -- normalized to $1 = \int L(v|\Model) \dd v$ -- the prediction of separate regions can be combined to realistic scenarios via convolution
\begin{eqnarray}
v_{\rm EG} & = & v_{\rm Local} + v_{\rm Host} + v_{\rm IGM} , \\
L_{\rm EG} & = & L_{\rm Local} * L_{\rm Host} * L_{\rm IGM} .
\label{eq:convolve}
\end{eqnarray}
This way, we predict the distribution $L(v_{\rm EG}|z)$ of the extragalactic component of the observed measures $v_{\rm obs}=v_{\rm MW} + v_{\rm EG}$ from FRBs at some redshift $z$, which can be compared to observations of localized FRBs with carefully subtracted Galactic foregrounds $v_{\rm MW}$.

In practice, the convolution of likelihoods is obtained by adding samples of identical size for each $L$ and computing the likelihood of the resulting sample.
The size of this sample is chosen to be the smallest size of samples used to compute individual $L$, usually $N \approx 5\cdot 10^4$ (see Sec. \ref{sec:IGM}).
The error of the convolution is given by the shot-noise of this sample. This is a more conservative estimate than following Gaussian error propagation of individual deviations.

For some regions, e.g. intervening galaxies (Sec. \ref{sec:inter}), the norm of $L$ is $<1$, representing the likelihood of no contribution.
For computation of the convolution, we consider an amount of $1-\rm norm$ of events in the corresponding sample to be equal to zero.

Some measures (e.g. RM) can have a positive or negative sign, allowing for contributions from different regions to cancel each other.
To account for that, each value in the sample of the logarithmic distribution is attributed a random algebraic sign.

\subparagraph{Likelihood of observation}

The majority of FRBs is not localised and the source redshift, $z$, is unknown.
However, by assuming a distribution of host redshifts, a prior $\pi(z)$, described in Sec.~\ref{sec:FRBpoppy}, we obtain the distribution of some measure, $v$, expected to be observed, 
\begin{equation}
L(v) = \int \pi(z) ~ L(v|z)~ \dd z.
\label{eq:tele}
\end{equation}
These predictions can be directly compared to observations.
In App.~\ref{sec:telescope_predictions} we show the expected contribution of individual models to the signal observed by several instruments.

\subparagraph{Multiple measures}

For the observation of an event with a single measure, $v$, the likelihood
for this to occur in a model $\Model$ is the corresponding value of the likelihood $L( v|\Model)$, obtained in Eq.~\eqref{eq:tele}.
However, when considering multiple measures $v_i$, e.g. DM and RM, from the same event, we have to account for their common redshift.
Instead of multiplying their individual likelihoods $L_0(v_0) \times L_1(v_1)$, as would be done for separate events, the likelihood of the second measure is thus factored into the integral in Eq.~\eqref{eq:tele},
\begin{equation}
L(\vec v) = \int \pi(z) ~ L_0( v_0|z)~ L_1( v_1|z)~ \dd z = \int \pi(z) ~ \prod\limits_i L_i( v_i|z)~   \dd z.
\label{eq:combined}
\end{equation}
This way we use the full information provided by an observation with measures $v_i$ instead of reducing it to a ratio of measures \citep[cf. e.g.][]{akahori2016,vazza2018frb,Piro2018}.

\subparagraph{Bayes factor}

The model likelihood computed for a single model does not hold any information on its own.
Instead, comparing the likelihoods of competing models allows to identify the best-fit candidates and to rule out less likely models.
The Bayes factor $\Bayes$ is defined as the ratio of the marginal likelihoods of two competing models \citep[e.g. ][]{imagine_whitepaper},
\begin{equation}
\Bayes(v|\Model_1,\Model_0) = \frac{L(v|\Model_1)}{L(v|\Model_0)} .
\label{eq:Bayes}
\end{equation}
It quantifies how the observation of $v$ changes our corroboration from model $\Model_0$ relative
to $\Model_1$. 
By comparing all models to the same baseline model $\Model_0$, comparison of $\Bayes$ is straight forward.
$\Bayes < 10^{-2}$, i. e. 100 times less likely, is usually considered decisive to rule out $\Model_1$ in favour of $\Model_0$ \citep{jeffreys1961}.

According to Bayes theorem,
\begin{equation}
P(\Model|v) \propto L(v|\Model) \pi(\Model) ,
\label{eq:Bayes-theorem}
\end{equation}
in order to arrive at the ratio of posteriors $P$, $\Bayes$ has to be multiplied by the ratio of priors $\pi$ of the models, which quantifies our knowledge due to other observational and theoretical constraints.
However, the results of our approximate Bayesian computation should only be used for identification of trends rather than model choices, which need to be confirmed by further analysis \citep[cf.][]{robert2011}.

%% file: models.tex

\section{Models}
\label{sec:models}

In this section, we explain how to quantify the contributions from the different regions along the LoS.
In our benchmark scenario, we assume FRBs to be produced around magnetars, hosted by a representative ensemble of host galaxies.
We consider contributions of the IGM
as well as a representative ensemble of intervening galaxies and their intersection probabilities. 
Finally, we consider the expected distribution of host redshifts for FRBs observed by different telescopes.


\subsection{Intergalactic medium}
\label{sec:IGM}

We estimate the contributions of the IGM using a constrained cosmological simulation that reproduces known structures of the local Universe, such as the Virgo, Centaurus and Coma clusters.
This simulation was produced using the cosmological magnetohydrodynamical code {\sc ENZO} \citep{Bryan2014} together with initial conditions obtained following \citet{Sorce2016}.
The simulation starts at redshift $z=60$ with an initial magnetic field, uniform in norm and direction, of one tenth of the maximal strength allowed by CMB observations of PLANCK \citep{Planck2014}, i. e. $B_0 \approx 0.1\rm~nG$ comoving.
Hence, this simulations is called \primordial.
Structure formation and dynamo amplification processes are computed up until redshift $z=0$, providing us with a realistic estimate of the structure of IGM and residual magnetic fields at high redshift as well as for the local Universe.
The constrained volume of $(250 ~{\rm Mpc}/h)^3$ that resembles the local Universe is embedded in a full simulated volume of $(500 ~{\rm Mpc}/h)^3$, in order to minimize artifacts from boundary conditions.
The adaptive mesh refinement applied in the central region allows to increase resolution in high-density regions by 5 levels to a minimum scale of $\approx 30 ~\rm kpc$.
Further information on this model can be found in \citet{hackstein2018} and \citet{hackstein2019}.
A reduced version of this model can be found on \href{https://crpropa.desy.de}{crpropa.desy.de} under \say{additional resources}, together with the other models probed in \citet{hackstein2018}.

\subparagraph{Probability estimate}

We extract the simulation data along different LoS, using the {\sc LightRay} function of the {\sc Trident} package \citep{hummels2017trident}.
This returns the raw simulation output of all physical fields within each cell of the LoS.
The distribution of results from all LoS to the same redshift $z$ is used to assess the likelihood of measures for FRBs hosted at this redshift, e.g. $L( \DMIGM | \zFRB)$.
With $\approx 50000$ LoS, likelihoods above 1  per cent have a shot noise below 0.05  per cent. 

\subparagraph{Cosmological data stacking}

The cosmological simulation provides snapshots at several redshifts, namely $z_{\rm snaps} \in [ 0.0,0.2,0.5,1.5,2.0,3.0,6.0 ]$.
To extract LoS, we stack the data \citep[e.g. ][]{Silva2000,akahori2016}.
As the path lengths of LoS within a redshift interval exceeds the constrained comoving volume of $($250 Mpc$/h)^3$, we combine randomly oriented segments until the redshift interval is completed.
The segments in a snapshot are computed for redshifts above $z_{\rm snaps}$.
This implies that the increased clumping of matter, expected at the end of a redshift interval, is also assumed for higher redshifts within the same interval, which may slightly over-estimate (within a factor $\lesssim 2$) the local matter clustering there, as well as the predictions for RM, SM and $\tau$.
However,  given that the DM is mostly due to IGM in voids, walls and filaments \citep[e.g. ][]{zhu2018scattering}, effects from over-estimation of matter clustering are negligible.

\subparagraph{Intergalactic DM}

\begin{figure}
    \centering
    \includegraphics[width=0.5\textwidth]{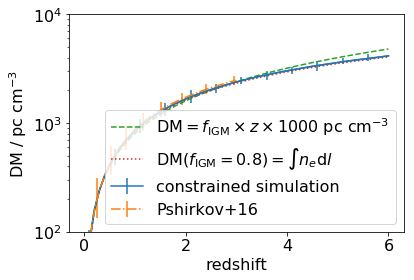}
    \includegraphics[width=0.5\textwidth]{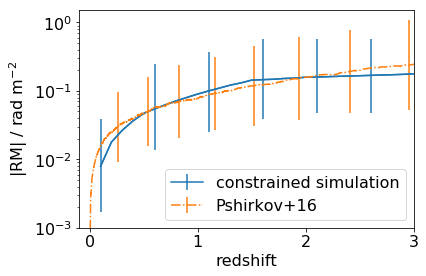}
    \caption{$\avg{\DMIGM}$ (top) and $\avg{\RMIGM}$ (bottom) as function of source redshift $z$ as obtained from IGM simulation (solid-blue) compared to parametrization (dashed-green \& dotted-red) and theoretical prediction obtained via Monte-Carlo simulation (dash-dot-orange).
        For consistent comparison to estimates following \citet{pshirkov2016}, we use $\fIGM \approx 0.83$, $l_c = 1 ~\rm Mpc$ and $B_0 = 0.1 ~\rm nG$.
    }
    \label{fig:DM_redshift}
\end{figure}
We obtain the proper free electron number density $n_e= \rho / (m_p ~\mu_e)$ from the gas density $\rho$ with proton mass $m_p$ and molecular weight of electrons $\mu_e = 1.16$, assuming that hydrogen and helium in the IGM are completely ionized, a common assumption after the epoch of reionization.
With this, we compute the DM along the LoS using
\begin{equation}
\DM(\zFRB) = \int\limits_0^{d(\zFRB)} n_e(z) ~(1+z)^{-1} ~\dd l(z) .
\end{equation}
The distribution of results along several LoS provides the expected likelihood of DM from sources at redshift $\zFRB$, $L(\DMIGM|\zFRB)$.
From this we can compute the estimated mean value 
\begin{equation}
\avg{\DM}(z) = \int \DM \times L(\DM|z) ~\dd\DM.
\end{equation}
The $\avg{\DM}$-redshift-relation obtained from the IGM simulation is in good agreement with \citep[cf.][]{Niino2018,connor2019}
\begin{equation}
\avg{\DM} =  z \times 1000 \rm ~ pc ~cm^{-3} ,
\label{eq:DM_IGM}
\end{equation}
as well as with the predictions obtained following \citet{pshirkov2016}.
For the latter, we perform a Monte-Carlo simulation, where we divide the LoS in segments of Jeans-length size and pick random $n_e$ from a log-normal distribution according to Eq. 2 in \citet{pshirkov2016}.
In Fig.~\ref{fig:DM_redshift} we compare these numerical expectations with theoretical expectations of DM for FRBs at cosmological distance with uniform IGM
\begin{equation}
\avg{\DM(z)} = \frac{c \rho_{\rm crit} \Omega_b \fIGM }{ m_p\mu_e H_0 }\int \frac{(1+z)}{H(z)} ~\dd z ,
\label{eq:DM_cosmo}
\end{equation}
with Hubble parameter $H(z)$ and $H_0 = H(z=0)$.
We use the critical density $\rho_{\rm crit}$ and baryon content of the Universe $\Omega_b$ from \cite{Planck2014}.

\subparagraph{IGM baryon content}

The results of our constrained simulation  agree well with Eqs. \eqref{eq:DM_IGM} and \eqref{eq:DM_cosmo} if we assume that a fraction $\fIGM = 1$ of baryons resides in the IGM.
This is expected as the limited resolution of the simulation does not allow us to properly resolve galaxy formation and the condensation of cold gas out of the IGM. 
However, it is estimated that in the local Universe about $18 \pm 4$  per cent of baryonic matter is in collapsed structures \citep{shull2012}.
The $7 \pm 2$  per cent of baryons found in galaxies are accounted for in the other models in Sec. \ref{sec:local} - \ref{sec:inter}.
In order to conserve the amount of baryons in our consideration, we have to subtract this part from $\fIGM$ and adjust results of our constrained simulation accordingly,
\begin{equation}
L(\DM|\fIGM) = \fIGM \times L(\fIGM\times\DM) .
\label{eq:shift_fIGM}
\end{equation}
\citet{pshirkov2016} assume $n_e = 1.8\times 10^{-7} ~\rm cm^{-3}$ at $z=0$, which implicitly assumes $\fIGM \approx 0.83$.
We choose this value to compute the other graphs in Fig. \ref{fig:DM_redshift}.

\subparagraph{Intergalactic RM}

\begin{figure}
    \includegraphics[width=0.5\textwidth]{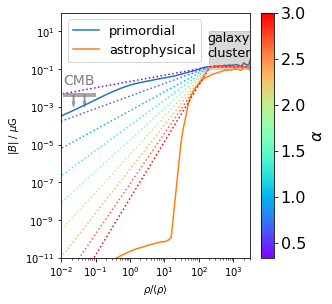}
    \caption{Median magnetic field strength $B$ as function of gas density $\rho$.
        The solid lines represent MHD simulations of extreme scenarios, i. e. primordial magnetic field of maximum allowed strength (\primordial, blue) or minimum strength (\astrophysical) together with astrophysical dynamo processes and magnetic feedback of AGN.
        We parametrize this shape with Eq.~\eqref{eq:B_rho}.
        The dotted lines represent different values of index $\alpha$, indicated by the colorbar.
        Constraints by \citet{Planck2015PMF} and \citet{vanWeeren2019} are indicated by the gray line and shade, respectively.            }
    \label{fig:rho-B_relations}
\end{figure}
The contribution to RM scales with the electron density times the magnetic field strength.
Accounting for cosmic expansion,
\begin{equation}
\RM(\zFRB) = \int\limits_0^{d(\zFRB)} B_\parallel ~n_e(z) ~(1+z)^{-2} ~\dd z .
\label{eq:RM_IGM}
\end{equation}
In Fig. \ref{fig:DM_redshift} we compare results to theoretical predictions obtained following \citet{pshirkov2016}.
The LoS magnetic field is obtained assuming that $B \propto n_e^{2/3}$, with a random change in direction after several Jeans lengths, which is assumed to be the coherence length.
We use a correlation length of $l_c = 1~\rm Mpc$ and $B_0 = 0.1 ~\rm nG$, in order to match the settings of the constrained simulation \primordial.
The results agree sufficiently well.
The estimates following \cite{pshirkov2016} assume a steeper $B\sim n_e$ relation than the constrained simulation, thus show slightly lower $\avg{\RM}$ for $0.5 \lesssim z \lesssim 2.0$, while the more realistic history of magnetic fields at higher redshift account for the decreased slope $z \gtrsim 1.5$.

Regardless of the magnetic field strength, low-density regions contribute very little to the observed signal, i.e.  $\ll 1$  per cent of $\RMIGM$,  making them hard to be detected and easily overshadowed by other regions along the LoS.
Hence, $\avg{RM}$ is not a direct measure of IGMFs in voids.
A more detailed discussion on this matter can be found in App. \ref{sec:RM_overdensity}.

Given the present lack of available observational detections of extragalactic magnetic fields beyond the scale of clusters of galaxies,  there is a large uncertainty in the strength of magnetic fields at higher over-densities, even up to $\rho/\avg{\rho} \approx 200$ \citep{vazza2017}.
While within galaxies and galaxy clusters magnetic fields are known to be $0.1 - 10 \rm ~ \mu G$ \citep{vanWeeren2019}, models for the origin and amplification of IGMFs differ in their predictions at intermediate density scales, $10 < \rho/\langle\rho\rangle < 200$ \citep[e.g. ][]{vazza2017}, associated with filaments and sheets, capable to imprint a detectable signal on $\avg{RM}$. 
Still, investigation of $L(\RM)$ is much more promising than $\avg{\RM}$, as it allows for a more detailed investigation of LoS crossing different regions of over-density.

\subparagraph{Model IGMFs}

By parametrizing the slope of the $B$-$\rho$-relation at lower densities, based on different simulations, we can evaluate different shapes and provide general constraints for models of the IGMF.  This allows us to quantify the likelihood for a variety of models based on a limited set of parameters without having to perform new simulations.

A simple parametrization is
\begin{equation}
|B| = \beta \rho^\alpha ,
\label{eq:B_rho}
\end{equation}
where we vary $\alpha$ and choose $\beta$ accordingly to match the simulated value at $\rho/\avg{\rho} = 200$.
The magnetic field-density relation for different $\alpha$ is shown in Fig. \ref{fig:rho-B_relations}.
\\

In order to estimate the LoS magnetic field $B_\parallel$ according to $\alpha$, we use the ratio of relations in Fig. \ref{fig:rho-B_relations} as renormalization factor for $B_\parallel$ extracted from \primordial, dependent on the local over-density $\rho/\avg{\rho} < 200$.
This procedure does reasonably well in reproducing the statistics of other  IGM simulations and allows for rapid investigation of an extensive set of magnetic models. However, we do not explore different magnetic field topologies this way.

Due to their overall similarity in the interesting  $1 \leq \rho/\avg{\rho} \leq 200$ range of density, as well as to minimise numerical artifacts, we identify the \primordial~  model with $\alpha = \frac{1}{3}$, which we use as the baseline to compute the renormalization factor for other choices of $\alpha$.
$\alpha=\frac{1}{3}$ is thus representative for the upper limit on IGMF strength provided by \citet{Planck2015PMF}, while $\alpha = \frac{9}{3}$ is representative for the lower limit on IGMF strength provided by \citet{neronov2010}.
The range of $\alpha$ thus roughly brackets all possible cases for the IGMF.

\subparagraph{Intergalactic scattering}

To compute the effective SM, as in \citet{zhu2018scattering}, we assume that turbulence in the IGM follows a Kolmogorov spectrum
\begin{equation}
\SMIGM \approx  
1.42 \cdot 10^{-13} ~ {\rm kpc~m^{-20/3}}
\ufrac{\Omega_b}{0.049}^2 \ufrac{L_0}{pc}^{-2/3} 
\label{Eq:SM_IGM}
\end{equation}
\begin{equation*}
\qquad \qquad \qquad \qquad \qquad 
\times 
\int\limits_0^{d} \bfrac{\rho(z)}{\avg{\rho}(z)}^2 (1+z)^4 \ufrac{\dd l}{kpc} .
\end{equation*}

\citet{macquart2013temporal} state that $L_0$ can lie between 0.001 pc and 0.1 Mpc.
\citet{zhu2018scattering} require $L_0 \approx 5 ~\rm pc$ in order to explain the $\tauIGM = 1 - 10 \rm~ ms$ scattering time at 1 GHz to be produced by the IGM alone.
However, according to \citet{lazio2008angular}, the large scales available in IGM would even allow for $L_0 \approx 1~\rm Mpc$.
\citet{Ryu_2008} investigate the IGM with hydrodynamical simulations and find that typical cosmological shocks during structure formation have curvature radii of the order of $\sim \rm~few~ Mpc$ and represent the characteristic scale of dominant eddies.
We adopt the latter as a reference here, and assume a constant $L_0 = 1 \rm~Mpc$ out to redshift 6. $L_0$ can be varied in post-processing, by applying a global factor, even if this is beyond the scope of this paper.
Still, for $L_0 = 1~\rm Mpc$, contributions of the IGM to temporal smearing $\tau$ are much lower than assumed in other work \citep[e.g.][]{zhu2018scattering}.
\\

The IGM is distributed along the entire LoS, barring negligible parts within host galaxy and MW.
Thus, to estimate $\tau$ (Eq.~\eqref{eq:tau}), $\Deff$ should be of the order of half the distance to the source, which would be the ideal position for a hypothetical lens \citep{macquart2004}.
For a possible source redshift $\zFRB$, we find the redshift $z_L$ of a hypothetical lens that maximizes
\begin{equation}
\Deff(\zFRB, z_L) = \frac{D_A(0,z_L)D_A(z_L,\zFRB) }{ D_A(0,\zFRB) } .
\end{equation}
We use the resulting values of $z_L$ and $\Deff$ in Eq.~\eqref{eq:tau} to calculate $\tau_{\rm IGM}$ from $\SMeff$ obtained for FRBs at redshift $\zFRB$.

In practice, it is not necessary to calculate $\tauIGM$ for each LoS individually.
Instead, Eq.~\eqref{eq:tau} implies identical shape of the likelihood for $\SMIGM$ and $\tauIGM$ for sources at redshift $\zFRB$, 
\begin{equation}
L(\tauIGM(\SMIGM,\zFRB)|\zFRB) \propto L(\SMIGM|\zFRB) ,
\label{eq:likelihood_transform}
\end{equation}
where the integral over $\tauIGM$ normalizes to 1.


\subsection{Host galaxies}
\label{sec:host}

In this section, we describe the model for density and magnetic fields in galaxies.

\subsubsection{Model description}

\citet{lacey2016unified} studied the evolution of galaxies with the semi-analytic galaxy formation model \galform. 
Dark matter halos in an N-body simulation provide a halo merger tree.
Furthermore, these halos provide a seed for individual galaxies, whose formation is modelled using differential equations for gas cooling, angular momentum and star formation.
Using the evolution of halo properties, including their merger history, they study the evolution of galaxies with a set of coupled differential equations of global galaxy parameters that correspond to well-defined astrophysical processes in galaxies, including AGN as well as stellar feedback.
\citet{lacey2016unified} provide a set of best-fit initial parameters for galaxy formation theory that reproduces the observed galaxy stellar mass function $\Phi(M_\star,z)$, morphological fractions, stellar metallicity, the Tully-Fisher relation as well as several luminosity functions.
The final output of the model is a large sample of galaxies that represents the expected ensemble 
of galaxies. For brevity, we will refer to set of time evolving properties of an galaxy in
\galform's output as a `galaxy model'.

The sample includes, both, central and satellite galaxies, where the latter corresponds to the most massive galaxy after a halo merger.
Since most stellar mass is concentrated in the more massive central galaxies, there is small likelihood for satellites to host FRBs.
For simplicity, we thus consider only central galaxies.

\subparagraph{Model galactic magnetic field}
\citet{rodrigues2018} use the results presented by \citet{lacey2016unified} with an optimised size-mass relation.
They investigate the evolution of magnetic fields for galaxies using the \magnetizer ~code  \citep{MagnetizerCode}, which numerically solves nonlinear turbulent mean-field dynamo theory \citep[e.g.][]{beck1994,arshakian2009,chamandy2014magnetic},  assuming thin galactic discs and axial symmetry.
For the small-scale magnetic field, they assume that the energy density is half of the interstellar turbulence energy density. 
This small-scale field serves as a seed field for the large-scale magnetic field and does not enter the computation of RM, for which we only use the coherent field component produced by the turbulent mean-field dynamo.
These equations deliver radial profiles of the strength of radial and toroidal components, while the axial component is obtained via $\nabla \cdot \vec{B} = 0$.
For the dependence on the axial coordinate, the magnetic field strength is assumed to be proportional to density, which declines exponentially.
This description of the coherent magnetic field allows to reconstruct the magnetic field along a LoS of arbitrary orientation and, together with the radial profile of free electron density, can be used to compute LoS integrals through the galaxy.

\subparagraph{Galaxy sample}

\citet{rodrigues2018} provide a sample of a few million galaxies, in agreement with current observations \citep[see][]{lacey2016unified}.
This sample represents the ensemble of galaxies in the Universe, thus prior expectations $\pi(\epsilon)$ for distribution of galaxy properties $\epsilon$, e.g. star formation rate, stellar population, metallicity, luminosity and circular velocity.
The total stellar mass $M_\star$ of these galaxies ranges from $10^7 ~M_\odot$ to $10^{12} ~M_\odot$.
By different combinations of disc and bulge properties, all morphologies of axisymmetric galaxies can be reproduced.
The sample thus includes spiral, lenticular and elliptical galaxies, represented by spherical galaxy models, but does not include irregular or peculiar galaxies, which account for only 
$\approx 5$ per cent of galaxies.

\subparagraph{Magnetic fields in \Rodrigues~ sample}

A prediction of \citet{rodrigues2018} is that a significant number of galaxies at $z=0$, especially with low $M_\star$, have very weak large-scale magnetic fields $<0.05 ~\mu\rm G$, because the conditions for a large-scale galactic dynamo are not satisfied.
They find evidence for their claim in a sample of 89 galaxies compiled by \citet{Beck2013}.

Furthermore, \citet{rodrigues2018} assume the large-scale field  to be destroyed completely by disc instabilities or during a merger of galaxies of comparable mass.
Hence, elliptical galaxies, that result from these processes, have weak regular magnetic fields.
Though, these fields can be amplified to $\mu$G strength in a time scale of $2-3$ Gyr \citep{arshakian2009},
estimates of RM for elliptical galaxies with vanishing coherent fields are mostly determined by numerical noise, since only the large-scale magnetic field enters computation.
They are, thus, too low to provide a significant contribution to observed RM. 
However, observations showed fluctuation of mainly low RM with amplitude of order $\lesssim 10\unitRM$ throughout elliptical galaxies \citep[e.g.][]{owen1990detection,clarke1992vla}.
Thus, ellipticals are expected to not contribute significantly to the observed RM. 
We hence argue that for the purpose of statistical investigation of measurable RM, the \Rodrigues~ sample is well-suited to represent the entire ensemble of galaxies.

\subsubsection{Probability estimate}

We obtain the likelihood $L(\DMHost|\zFRB)$ for the contribution of an unknown host at redshift $\zFRB$ by a prior weighed integral, e.g.
\begin{equation}
L(\DMHost|z) =  \int  L'(\DMHost|\epsilon,z) ~\pi(\epsilon|z) ~\dd \epsilon ,
\end{equation}
where $L(\DMHost|\epsilon,z)$ is the likelihood of the expected contribution for an individual galaxy with properties $\epsilon$ at redshift $z$.
The prior of $\epsilon$ at $z$ is denoted by $\pi(\epsilon|z)$.
\begin{equation}
L'(\DM|\epsilon,z) = (1+z) L((1+z) DM|\epsilon)
\end{equation}
is the likelihood of the signal as seen by the observer, computed from the modeled expectation of residual DM. 
Similar relations hold for SM and RM, that evolve as $(1+z)^{-2}$ \cite[cf. e.g. ][]{hackstein2019}.

\subparagraph{Monte-Carlo simulation}

In practice, it is not necessary to compute the full likelihood $L(\DM|\epsilon)$ for each galaxy model.
Instead, we do a Monte-Carlo experiment and repeatedly pick a random axisymmetric model, inclination angle and impact parameter.
By choosing the sample according to priors $\pi$, the distribution of those results provides us with the required likelihood.

For the impact parameter, we naturally assume uniform $\pi$, while the inclination is sampled from a cosine distribution, expecting more galaxies face on, according to the orientation of galaxies in the local supercluster \citep[e.g.][]{hu1995,yuan1997}.
We assume FRBs to be produced by magnetars, which are most likely found in the vicinity of star-forming regions.
Molecular gas, which allows for effective cooling, is a good tracer of star-forming regions \citep[e.g.][]{arce2006molecular}.
Along the LoS, defined by inclination and impact parameter, we compute the integral to position of the source.
The path integral is computed only within an ellipsoid whose major axis and disc size, respectively, are 3.5 times scale height and 2.7 times half-mass radius of a given galaxy model, which marks the distance where surface mass density reaches 1 per cent of the central value.
The position of the source is picked randomly according to the profile of molecular gas density.
The LoS is excluded, if it does not enter the galactic ellipsoid or in case that the molecular gas density along the LoS does not surpass a minimum value of $\rho_{\rm mol} \gtrsim 10^{-37} ~\rm g ~cm^{-3}$, chosen for numerical reasons, which is too low to indicate the possible habitat of magnetar FRB progenitors.
For this choice, the models are representative for the contribution of stellar disks of galaxies.
However, a physically motivated certainly higher limit on $\rho_{\rm mol}$ would even more concentrate the assumed distribution of source positions on the dense part of their host galaxies and thus account for increased contributions to the observed measures.

Furthermore, the likelihood for LoS to contain a possible FRB progenitor is proportional to the column density of molecular gas.
However we argue that this is dominated by path length and the resulting likelihood function is well reproduced by disregarding LoS with probability given by the path length through the ellipsoid (see App. \ref{sec:HostPathLength}) divided by maximum path length, i. e. disk diameter.
\\

The galaxy population modelled by \cite{rodrigues2018} represents theoretical prior expectations $\pi(\epsilon|\zFRB)$ of the distribution of galaxy properties $\epsilon$ at different source redshifts $\zFRB$.
Sampling the entirety of this population naturally accounts for this prior assuming that all types of axisymmetric galaxies host FRBs.
However, more massive galaxies contain a greater number of stars, thus are more likely to host FRB progenitors.
To account for this, we multiply the prior of galaxies by their total stellar mass $M_\star$.
We pick a sample of $\approx 10^6$ galaxy models and compute for each a number of 10 LoS.
The results for this sample of $\approx 10^7$ LoS provides a converged estimate of the likelihood for the host contribution, without knowledge of the inclination angle, source position or galaxy type. With this sample, likelihoods above 1 per cent are accurate to less than 0.003 per cent.

\subsubsection{Host scattering}

To estimate the SM contributed by the host galaxy, we use Eqs. \eqref{eq:SM} \& \eqref{eq:c_ngal}. We set $L_0 = 0.1 ~\rm kpc$ to the maximum size of supernova remnants, before they drop below the sound speed.

Finally, we calculate $\tau$ from Eq.~\eqref{eq:tau}.
Obviously, $z_L$ is identified with the redshift of the host galaxy $\zFRB$.
Inside the host galaxy, the angular diameter distance to source and plasma along the LoS are almost identical, $D_S \approx D_L$.
hence $\Deff \approx D_{LS}$.
To estimate scattering in the host galaxy, $\Deff$ should be characteristic for the distance to the bulk of material \citep{macquart2004}.
A reasonable choice is half the path length of LoS inside the host galaxy, obtained for the individual LoS.
The same choice is a fair approximation for scattering in the MW.
Here, we approximate the path length by the redshift-dependent average size of galaxies of the probed sample.
This assumption yields a reasonable estimate on the magnitude of $\tau$, which is below $< 10 ~\rm ns$, hence not observable by current instruments.


\subsection{Intervening galaxies}
\label{sec:inter}

\subparagraph{Model description}

The results of \citet{rodrigues2018}, used to model the host galaxy in Sec. \ref{sec:host}, can also be used to model the contribution of intervening galaxies.
The expectation for a variety of galaxies can be computed in the same manner, i.e. for a random inclination angle and impact parameter we can compute LoS integrals through the entire galaxy.
By sampling the galaxy population of \citet{rodrigues2018} at redshift $\zInter$, we obtain the model likelihood for contributions from intervening galaxies at this redshift, $L(\RMInter|\zInter)$.
Of course, the impact parameter and the inclination angle have a prior with uniform and cosine shape, respectively (cf. Sec. \ref{sec:host}).
However, we only consider LoS within the ellipsoid representing the galaxy model, which is considered to where it falls below 1 per cent of the central surface mass density (cf. Sec. \ref{sec:host}).
Smaller galaxies have less chance to intersect a LoS and in order to account for this, we multiply the prior of galaxies by their squared half-mass radius.

\subparagraph{Intersection probability}

\begin{figure}
    \centering
    \includegraphics[width=0.4\textwidth]{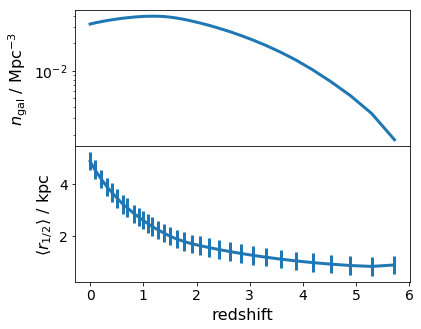}
    \includegraphics[width=0.4\textwidth]{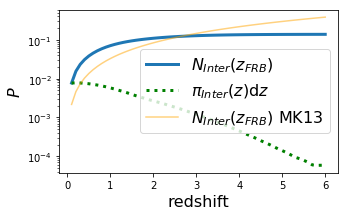}
    \caption{ Top: comoving number density $n_{\rm gal}$ and average half-mass radius $r_{1/2}$ of the considered galaxy sample as function of redshift.
        Galaxies grow in average size ($\dd r_{1/2} / \dd z <0$), mostly due to expansion and mergers, which also reduces their number in a fixed volume ($\dd n_{\rm gal} / \dd z <0$), together determining the shape of $\piInter(z)$.
        Note that we do not consider galaxies with stellar mass $M_\star < 10^7 M_\odot$, causing $n_{\rm gal}$ to go down at high redshift.
        The implicit number density assumed for galaxies with different mass threshold are considered according to galaxy stellar mass function and redshift evolution \citep[see][]{lacey2016unified}.
        Bottom: average number of intervening galaxies (solid blue) in LoS to source at redshift $\zFRB$ and prior (dotted green) for intervening galaxy at $z$ (Eq.~\eqref{eq:N_inter}) per redshift interval $\dd z = 0.1$.
        The thin orange line shows the expectation of \citet{macquart2013temporal}.
    }
    \label{fig:Ninter}
\end{figure}
The mean number of intersecting galaxies along a LoS to source at redshift $\zFRB$ can be estimated by \citep{macquart2013temporal}
\begin{equation}
\NInter (\zFRB) = \int\limits_0^{\zFRB} \pi r_{\rm gal}^2 ~n_{\rm gal} \frac{d_H(z)}{(1+z)} ~\dd z = \int\limits_0^{\zFRB} \piInter(z)~ \dd z,
\label{eq:N_inter}
\end{equation}
with galaxy radius $r_{gal}$, galaxy number density $n_{gal}$ and Hubble radius $d_H(z)$. 
By definition, the complementary cumulative galaxy stellar mass function yields the number density of galaxies as function of minimum mass $M_0$
\begin{equation}
n_{\rm gal}(>M_0, z) = \int\limits_{M_0}^{\infty} \Phi(M_\star,z) ~\dd M_\star .
\label{eq:n_gal}
\end{equation}
By  accounting for $\Phi(M_\star,z))$ in the \Rodrigues~ sample \citep[see][]{lacey2016unified}, we obtain realistic contribution from intervening galaxies of all $M_\star>M_0$, independent on the choice of $M_0$.

We obtain $n_{\rm gal}$ from the number of galaxies and considered volume of the \Rodrigues~ sample and $\avg{r_{\rm gal}}$ as 2.7 times the average half-mass radius of galaxy models used to sample $L$ (cf. Sec. \ref{sec:host}).
Thus, $\avg{r_{\rm gal}}$ considers the galaxies weighted by their intersection probability $\propto r_{\rm gal}^2$.
In Fig. \ref{fig:Ninter}, top, we show both, $n_{\rm gal}$ and $\avg{r_{\rm gal}}$, as function of redshift.
Galaxies increase their mass and volume over time, thus $\avg{r_{\rm gal}}$ decreases with redshift.
Mergers also reduce the number of galaxies within a fixed volume, thus $\frac{\dd n_{\rm gal}}{\dd z }>0$.
However, we only consider galaxies with $M_\star > 10^7 M_\odot$, which have to grow from smaller galaxies at higher redshift that we do not account for.
Thus, $n_{\rm gal}$ decreases at high redshift.
However, since we consider all galaxies $> 10^7 M_\odot$, independent of a brightness limit, $n_{\rm gal}\approx 0.03 ~\rm Mpc^{-3}$ at $z=0$, significantly higher than assumed elsewhere \citep[e.g. $n_{\rm gal}\approx 0.007 ~\rm Mpc^{-3}$ in][]{macquart2013temporal}.

\subparagraph{Probability estimate}

The integrand of Eq.~\eqref{eq:N_inter} defines a prior $\piInter(z)$ for the LoS to intersect a galaxy at redshift $z$, which can be used to obtain the likelihood, e.g. of $\RM$, from intervening galaxies along a LoS to source redshift $\zFRB$,
\begin{equation}
L(\RMInter|\zFRB) = \int\limits_{0}^{\zFRB} L(\RMInter|z)~ \piInter(z) ~ \dd z.
\end{equation}
By sampling the entire ensemble of models provided by \citet{rodrigues2018}, all types of axisymmetric galaxies could intervene the LoS.
We pick a sample of $\approx 10^6$ galaxy models and compute for each a number of 10 LoS.
The results for this sample of $\approx 10^7$ LoS provides a converged estimate of the likelihood for the contribution of intevening galaxies, without knowledge of the inclination angle, galaxy type or position along the LoS.
Again, likelihoods above 1  per cent are accurate to less than 0.003  per cent.

\subparagraph{Probability of intervening galaxies}

We assume that all progenitors of FRBs are located within a galaxy.
Thus, for a FRB hosted at redshift $\zFRB$, the normalization, e. g. of $\int L(\RMHost|\zFRB) ~\dd \RMHost = 1$, indicates that the host contributes $\RMHost$ within the range of $L(\RMHost|\zFRB)$ to each LoS.
In order to represent the probability of intersecting another galaxy, $L(\RMInter|\zFRB)$ must be normalized to the expected average number of intervening galaxies per LoS, (Eq. \eqref{eq:N_inter})
\begin{equation}
\NInter (\zFRB) = \int L(\RMInter|\zFRB) ~\dd \RMInter ,
\label{eq:Ninter_renorm}
\end{equation}
indicating that $\RMInter$ are only contributed to $\NInter < 100$ per cent of LoS.
The correct normalization $\NInter$ highly depends on the choice of $r_{\rm gal}$, which should thus represent the size of galaxy model considered for computation.
The results for $\NInter(z)$ and $\piInter(z)$ are shown in Fig. \ref{fig:Ninter}.
Compared to results of \citet{macquart2013temporal}, we expect more galaxies to intersect the LoS to low redshifts $z<3$, e.g. they expect less than 5  per cent of LoS to $z=1.5$ compared to $<10$ per cent for the \Rodrigues~ sample, which is due to the $\approx 4$ times higher $n_{\rm gal}$ at $z=0$.
However, we expect less LoS to high redshift $z>3$ to be intervened, e.g. they expect $< 40$ per cent for $z=4$, while only $<30$ per cent in \Rodrigues.
Though the decreasing size of galaxies is partly responsible, this feature is dominated by the artificial choice to not account for galaxies with $M_\star < 10^7 ~M_\odot$.

\subparagraph{Intervening galaxy scattering}

For the temporal smearing $\tau$ by an intervening galaxy, $\Deff$ depends on redshift of both, the source $\zFRB$ and the intervening galaxy $\zInter$, requiring explicit computation of $\Deff$ in Eq.~\eqref{eq:tau}, 
Since only global factors are applied to $\SMeff$, the expected contribution of intervening galaxies at redshift $\zInter$ to SM, $L(\SMInter|\zInter)$, and to $\tau$, $L(\tauInter|\zFRB,\zInter)$, observed for FRBs hosted at redshift $\zFRB$, are of identical shape (cf. Eq.~\eqref{eq:likelihood_transform}).
The likelihood $L(\tauInter|\zFRB)$ for contribution of an intervening galaxy at unknown redshift to the signal from source at $\zFRB$ is obtained by the prior-weighed integral over $\zInter$,
\begin{equation}
L(\tauInter|\zFRB) = \int L(\tauInter|\zFRB,z) ~\piInter(z)\,\dd z\,,
\end{equation}
with $\piInter(z)$ from Eq.~\eqref{eq:N_inter}.


\subsection{Local environment}
\label{sec:local}

\subparagraph{Model description and probability estimate}

Here, we assume that all FRBs are produced by magnetars \citep{metzger2017,zanazzi2020}.
The contribution to the DM and RM from the local environment of a young neutron star are described in \citet{Piro2018}.
More details on this model, the Monte-Carlo simulation to obtain probability estimates as well as the considered priors can be found in \citet{hackstein2019}, where we quantify predictions of the DM and RM.
We consider a sample of $10^7$ events, thus likelihoods above 1  per cent are accurate to less than 0.003  per cent.
Note that the majority of magnetars in this model are of decent age $>10^2 \rm~ yr$ and thus contribute rather low amounts of $\DM$ and $\RM$ \citep[cf. Figs. 7 \& 8 in][]{Piro2018}.

\subparagraph{Local scattering}

To estimate the SM contributed by the local environment of a magnetar, we use Eqs. \eqref{eq:SM} \& \eqref{eq:c_ngal}.
Calculation of the SM is hence almost identical to DM,
\begin{equation}
\SM = \alpha_c L_0^{-2/3} \int n_e^2 ~\dd l ,
\end{equation}
where $\alpha_c$ is a factor and $L_0$ the outer scale.
Assuming that $n_e$ is constant within the different regions of the supernova remnant, their contribution can be computed as \citep[cf. to Eqs. 10 \& 13 in][]{Piro2018}
\begin{align}
\SM_{\rm SNR} & =  \alpha_c L_0^{-2/3} n_r^2 (R_c - R_r) , \\
\SM_{\rm ISM} & =  16 \alpha_c  L_0^{-2/3} n^2  (R_b - R_c),
\end{align}
for the \textit{uniform} case and \citep[cf. to Eqs. 38 \& 39 in][]{Piro2018}
\begin{align}
\SM_{\rm SNR} & =  \alpha_c L_0^{-2/3} n_r^2 (R_c - R_r) , \\
\SM_{\rm w,sh} & =  16 \alpha_c  L_0^{-2/3} n^2  (R_b - R_c), \\
\SM_{\rm w,unsh} & =  \alpha_c  L_0^{-2/3} n^2  R_b,
\end{align}
for the \textit{wind} case, where $\alpha_c = 0.18 \rm~kpc~ m^{-20/3}$, $L_0$ is in pc, $n$ and $n_r$ are in $\rm cm^{-3}$.
Equations for $n_r$ as well as radii $R_b$ and $R_c$ are given in \citet{Piro2018}.
For $L_0$ we assume the size of the supernova remnant $R_b$.
To obtain the observed $SM_{\rm eff}$ caused by the local environment at cosmological distance, these results are shifted to the redshift, $\zFRB$, by applying factor $(1+\zFRB)^{-2}$, according to Eq.~\eqref{eq:SMeff}.
For our benchmark model, we consider magnetars in the {\it wind} case, embedded in an environment dominated by stellar winds from the heavy progenitor star.
\\

Inside the host galaxy, the angular diameter distance to source and lensing material are almost identical, $D_S \approx D_L$ (cf. Eq.~\eqref{eq:tau}),
hence $\Deff \approx D_{LS}$.
To estimate scattering in the host galaxy, $\Deff$ should be characteristic for the distance to the bulk of material \citep{macquart2004}.
A reasonable choice is half the path length of LoS inside the host galaxy, obtained for the individual LoS.
For the local environment of the FRB progenitor, $\Deff$ is well approximated by half the size of the environment.
In case of the magnetar model, this is half the size of the supernova remnant, $\Deff = R_b/2$.
Obviously, $z_L$ is identified with the redshift of the host galaxy $\zFRB$, allowing us to calulate $\tau$ from Eq.~\eqref{eq:tau}.


\subsection{Redshift distribution}
\label{sec:FRBpoppy}
\begin{figure}
    \centering
    \includegraphics[width=0.33\textwidth]{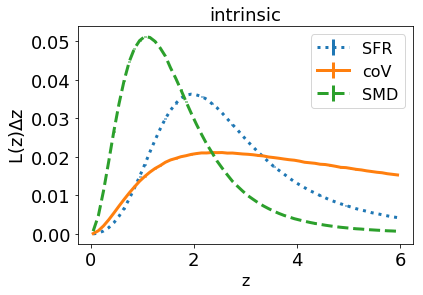}
    \includegraphics[width=0.33\textwidth]{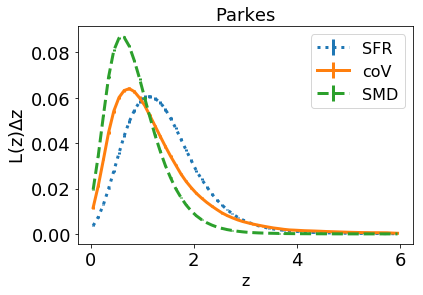}
    \includegraphics[width=0.33\textwidth]{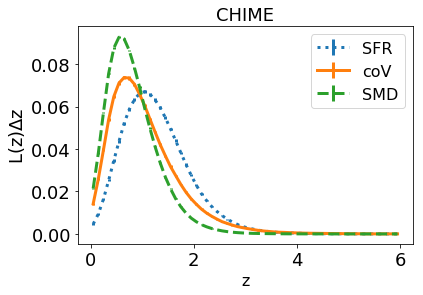}
    \includegraphics[width=0.33\textwidth]{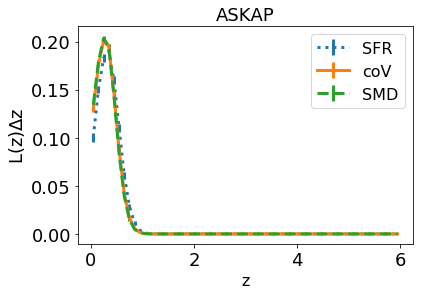}
    \caption{Top: Intrinsic distribution of host redshift for FRBs in case of FRB redshift distribution following stellar mass density (SMD, dashed), comoving volume (coV, solid) or star formation rate (SFR, dotted) (Eqs. \eqref{eq:population_coV} - \eqref{eq:population_SFR}).
        Others: distribution of redshifts, expected to be observed by Parkes, CHIME or ASKAP (top to bottom).
        These estimates serve as a prior for redshift $\pi(z)$ in the interpretation of $z$-dependent measures of unlocalized FRBs.
        The barely visible error bars show the shot noise of the Monte-Carlo sample. 
        The redshift bins are scaled linearily, thus each bin has the same $\Delta z = 0.1$.
        }
    \label{fig:population}
\end{figure}

Reasonable choices for the redshift prior of FRBs $\pi(z)$ should assume a physically motivated intrinsic distribution of $z$ and consider instrument responses that determine the detectable subset of the population.
{\sc frbpoppy}\footnote{\href{https://github.com/davidgardenier/frbpoppy}{github.com/davidgardenier/frbpoppy}} \citep{gardenier2019synthesizing} is a python-package built to investigate the population of FRBs.
It allows to assume reasonable intrinsic redshift distributions and to apply the selection effects of individual instruments due to sensitivity, wavelength range, or time resolution.

\subparagraph{Assumed intrinsic redshift distribution}

We consider three different intrinsic redshift distributions for FRBs, presented by \citet{gardenier2019synthesizing}.
The simplest assumption is a constant number density of FRBs, 
\begin{equation}
n_{\rm FRB} = \rm const.
\label{eq:population_coV}
\end{equation}
This suggests the redshift distribution of FRBs to have a
constant comoving density across epochs (coV).

Many models consider stellar objects or the merger of those as sources of FRBs.
These are more likely to occur in regions with a high number density of stars, thus suggesting the redshift distribution of FRBs to follow the evolution of the stellar mass density \citep[SMD,][]{madau2014},
\begin{equation}
n_{\rm FRB} = \int_z^\infty \frac{ (1+z')^{1.7} }{ 1+[1+z')/2.9]^{5.6} } \frac{\dd z'}{H(z')}  .
\label{eq:population_SMD}
\end{equation}

Young neutron stars and magnetars are widely considered to be the most likely sources of FRBs.
Such stars are more likely to be found in the vicinity of star-forming regions, implying the FRB redshift distribution to follow the evolution of the cosmic star formation rate  \citep[SFR,][]{madau2014},
\begin{equation}
n_{\rm FRB} = \frac{ (1+z')^{2.7} }{ 1+[1+z')/2.9]^{5.6} }.
\label{eq:population_SFR}
\end{equation}

All other parameters are set to the values of the \textit{complex} population presented in \citet{gardenier2019synthesizing}.
In Fig. \ref{fig:population} we show the intrinsic distribution of host redshifts, assuming the FRB population to follow SMD, coV or SFR, as well as corresponding $\pi(z)$  expected to be observed with ASKAP (in coherent mode), CHIME or Parkes.

\subparagraph{Probability estimate}

Using {\sc frbpoppy}, we generate a random sample of $10^7$ FRBs and their intrinsic properties, such as luminosity and pulse width, following one of the assumed redshift distributions.
Subsequently, we apply the selection effects of ASKAP, CHIME and Parkes to filter out FRBs that can actually be measured by those instruments.
The initial parameters are optimized in order to reproduce the observed distribution of DM and fluence
\citep[for more details, see][]{gardenier2019synthesizing}.
The redshift distribution of the intrinsic and selected samples is shown in Fig. \ref{fig:population}.
The latter serve as prior $\pi(z)$ on the host redshift of unlocalized FRBs observed by the corresponding telescope.
With a remaining sample size of at least $3 \times 10^{4}$, likelihoods above 1  per cent are accurate to less than $\lesssim 0.05$ per cent.

\subparagraph{Discussion} 

The main parameter responsible for the difference in source selection is the gain of the telescope.
The values of gain used in {\sc frbpoppy} ranges from $0.1~\rm K ~Jy^{-1}$ (ASKAP) over $0.69  ~\rm K ~Jy^{-1}$ (Parkes) to $1.4  ~\rm K ~Jy^{-1}$ (CHIME).
Since FRBs at large redshifts are too faint to be observed, our results suggest that the cosmic volume probed by ASKAP is not expected to go beyond $z \approx 1.0$.
In this range, the populations can hardly be distinguished since they are all dominated by the increasing volume.
However, Parkes and CHIME have rather similar $\pi(z)$ and the chance to observe FRBs at higher redshift $z>1.0$ differs reasonably between the assumed intrinsic redshift distributions.
The generally low distance of FRBs observed by ASKAP makes them more vulnerable to the unknown local contributions.
\\

Note that {\sc FRBpoppy} uses estimates, e.g. of $\DM(z)$, in order to decide how many FRBs will be observed at a given redshift.
Theses estimates have been produced using slightly different assumptions on the contributing regions.
However, the $\DM$ is dominated by the IGM and the analytical description used in {\sc FRBpoppy} provides a good match to our estimates.
Hence, we argue that this does not alter the general conclusions of this work.
In the future, we plan to converge the assumptions used in {\sc FRBpoppy} and {\sc PreFRBLE} in order to provide consistent results.
However, this is not trivial, as the change in one parameter can influence the best-fitting choice for other parameters, thus requires a repetition of the inference presented in \citet{gardenier2019synthesizing}.

%% file: applications.tex

\section{Applications}
\label{sec:applications}


\subsection{Identification of intervening galaxies}
\label{sec:identify}
\begin{figure*}
    \centering
    \includegraphics[width=\textwidth]{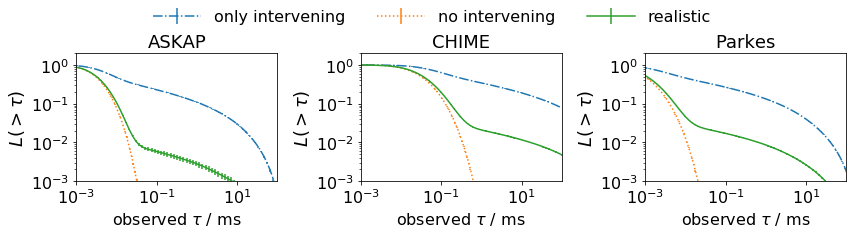}
    \caption{Complementary cumulative distribution of $\tau$ expected to be observed with ASKAP (left), CHIME (center) and Parkes (right) in our benchmark scenario, considering LoS with exactly one intervening galaxy (dotted-orange) or without any (dash-dotted-blue).
        The excess of the former at $\tau_0$ shows how many more FRBs are expected with $\tau>\tau_0$ for LoS with intervening galaxies.
        The solid green line shows expectations for a realistic mix of LoS with and without intevening galaxies.
    }
    \label{fig:tele_tau}
\end{figure*}

\subsubsection{Method}

In Fig.~\ref{fig:tele_tau} we show the complementary cumulative likelihood $L(>\tau)$ of extragalactic $\tau$ expected to be observed by different instruments in three similar versions of our benchmark scenario.
Each considers contributions from the local environment of the source, assumed to be a magnetar, the host galaxies and the IGM (see Sec.~\ref{sec:models}).
The three versions are
\begin{description}
\item[i. \emph{no intervening}] only LoS without intervening galaxies.
\item[ii. \emph{only intervening}] LoS with a single galaxy along the LoS, at random redshift according to $\piInter(z)$ (Fig. \ref{fig:Ninter}).
\item[iii. \emph{realistic}] LoS with and without intervening galaxies.
The ratio of their number for sources at redshift $\zFRB$ is given by $\NInter(\zFRB)$ (Fig. \ref{fig:Ninter}). 
\end{description}

We quantify the likelihood of FRBs observed with $\tau$ to have an intervening galaxy along the LoS by computing the Bayes-factor $\Bayes$ (Eq.~\eqref{eq:Bayes}) as the ratio of $L(\tau)$ in the two extreme scenarios. $\Bayes(\tau) > 100$ signals that $\tau$ is 100 times more likely to be observed in case of an intervening galaxy.
However, according to Bayes theorem (Eq.~\eqref{eq:Bayes-theorem}), in order to factor in our previous knowledge, $\Bayes$ has to be multiplied by the ratio of priors, which can be identified as the expected number of LoS which contain at least one intervening galaxy $\pi_{\rm I}$.
In our model, this can be obtained by integrating the expected number of LoS with intervening galaxies $\NInter(\zFRB)$ (Eq.~\eqref{eq:N_inter}) as function of source redshift $\zFRB$, multiplied by prior of source redshift $\pi(\zFRB)$, obtained in Sec. \ref{sec:FRBpoppy},
\begin{equation}
\pi_{\rm I} = \int \NInter(\zFRB)~\pi(\zFRB) ~\dd \zFRB .
\end{equation}
Assuming the intrinsic distribution of $\zFRB$ to follow SMD, we predict intervening galaxies along LoS for $\pi_{\rm I} = 2.5$  per cent, 5.9  per cent and 6.2  per cent of FRBs observed by ASKAP, CHIME and Parkes, respectively.

Multiplying the corresponding ratio of priors $\pi_{\rm I} / (1-\pi_{\rm I})$ to $\Bayes$ yields the ratio of posteriors $P$ (Eq. \eqref{eq:Bayes-theorem}).
However, the ratio of posteriors does not exceed $100$, marking 99 per cent certainty of an intervening galaxy along the LoS.
This is because the scenario without intervening galaxies cannot provide $\tau > 0.06 ~\rm ms$, according to our models, while the ratio of $P$ for slightly lower values of $\tau$ does not yet reach 100.

\subsubsection{Results}

For FRBs observed by ASKAP and Parkes at $\nu~\!=~\!1300\!~\rm MHz$, $\tau_{\rm dist} = 0.06\,\rm ms$ marks the minimum temporal broadening that is certainly associated to an intervening galaxy. 
Also, for FRBs observed by CHIME at lower characteristic frequency, $\nu = 600 ~\rm MHz$, where scattering effects are more severe (see Sec. \ref{sec:tau}), $\tau_{\rm dist} = 1.8 ~\rm ms$. 
We find that 26.8 per cent, 30.8 per cent and 30.6 per cent of the sightlines with intervening galaxies will show $\tau > \tau_{\rm dist}$, for ASKAP, Parkes and CHIME, respectively.
Thus, we predict that these telescopes observe 0.7 per cent, 1.9 per cent, and 1.8 per cent of FRBs with $\tau \geq \tau_{\rm dist}$.
However, for the FRBs listed in FRBcat, we find 3.6 per cent, 48 per cent and 20 per cent above the corresponding $\tau_{\rm dist}$.

\subsubsection{Discussion}

The expected number of LoS with intervening galaxies is smaller for ASKAP since
a narrower redshift range is probed than by CHIME and Parkes
(cf. Fig. \ref{fig:population}).
$\Deff$ is significantly smaller at $z<1$ and galaxies are denser and more turbulent at higher $z$, thus providing smaller $\tau$ at lower redshift.
The majority of LoS with $\tau < \tau_{\rm dist}$ either cross
smaller galaxies with a low contribution to all measures,
intersect only small parts of an intervening galaxy, or the additional galaxy is located close to source or observer, resulting in a sub-optimal $\Deff$.
Even though most of significant contribution to the other measures, i.e. DM and RM, will arise from the latter subset,
consideration of intervening galaxies is still necessary for reasonable interpretation of those measures.
\\

For all telescopes, the observed number of $\tau > \tau_{\rm dist}$ in FRBcat is 5 to 25 times more than expected.
Moreover, the total number of LoS with intervening galaxies is reasonably smaller than this number.
Thus, the high number of $\tau > \tau_{\rm dist}$ observed by Parkes can hardly be attributed to intervening galaxies alone, which might only account for $\lesssim 13$ per cent of these events.
This is despite the fact, that we expect a higher number of intervening galaxies than earlier works \citep[e.g.][]{macquart2013temporal}.
Note that we do not consider the circumgalactic medium, which would certainly increase
this estimate.

For the contribution of the IGM, we assume a physically motivated $L_0 = 1 ~\rm Mpc$, hence low contribution to $\tau$.
Still, in order for the IGM to account for the remaining events, $L_0 \lesssim 1 ~\rm pc$ would be required.

Our magnetar model for the environment local to the source  is the only region that provides $\tau \lesssim \tau_{\rm dist}$ (see App. \ref{sec:telescope_predictions}).
However, from the recent observation of an FRB-like radio burst from a Galactic magnetar, \citet{margalit2020implications} conclude that magnetars responsible for cosmological FRBs result from other origins than normal core-collapse supernovae, such as superluminous supernovae, accretion-induced collapses or neutron star mergers.
Such sources can produce visible FRBs somewhat earlier \citep{metzger2017,margalit2019fast}, in a much denser and more turbulent state of the remnant.
These models might thus account for a stronger scattering than our model.
\\
Considering a higher mass threshold for galaxies than $M_\star \geq 10^7 ~M_\odot$ will likely not affect the number of LoS observed with $\tau > \tau_{\rm dist}$ in the realistic sample of FRBs, with and without intervening galaxies.
This is because massive galaxies dominate $\tau$ and our model realistically considers the galaxy stellar mass function, thus the amount of galaxies with high mass, independent of the chosen minimum mass of small galaxies.
Still, other versions of galaxy formation theory might differ in their predictions, e.g. of turbulence in galaxies at large distance, thus potentially provide higher amounts of $\tau > \tau_{\rm dist}$, which will be visible in $L(\tau)$.
\\

Here we assume that the number density of galaxies $n_{\rm gal}$ is uniform in space.
However, $n_{\rm gal}$ increases with the gas density, as more galaxies reside in the dense environment of galaxy clusters.
Hence, a more sophisticated approach should consider clustering, e.g. via density profile of LoS, providing each with an individual prior for redshift of galaxy intersection, $\piInter(z)$.
This way, LoS with high contribution from IGM, associated with high-density regions, would have a higher chance of additional signal by intervening galaxies with an increased chance for multiple intersections.
In turn, for LoS that mainly traverse low-density regions, the chance for intervening galaxies would be lower.
Accounting for clustering of galaxies would increase the significance of results from RM of FRBs regarding IGMFs and their cosmic origin (Sec. \ref{sec:IGMF}).
However, in this work we are mostly interested in FRBs from high redshift, 
$z \gtrsim 0.5$, which are most indicative of the IGMF.
On this scale, the structure of the Universe can reasonably be considered as fairly homogeneous.
We argue that for FRBs from high redshift the statistical results are almost identical to the more sophisticated approach, which is necessary only for the correct interpretation of FRBs from lower redshift.

Note that it is possible to obtain an estimate on the redshift of an intervening galaxy, $\zInter$, by comparing scenarios with $\piInter(z) = \delta(z-\zInter)$ for different possible $\zInter$.
This is, however, beyond the scope of this paper and will be investigated in the future.


\subsection{Redshift estimate}
\label{sec:redshift}
\begin{figure*}
    \centering
    \includegraphics[width=0.45\textwidth]{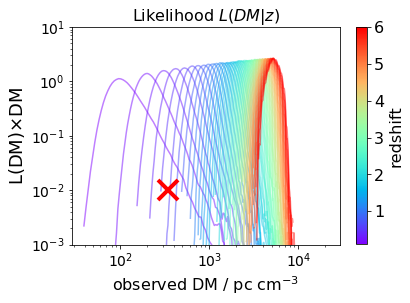}
    \includegraphics[width=0.45\textwidth]{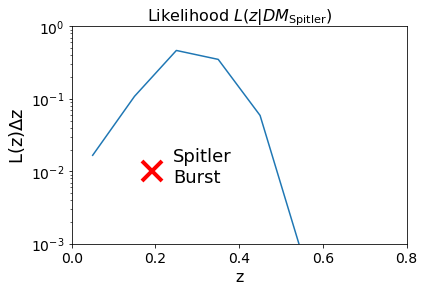}
    \includegraphics[width=0.45\textwidth]{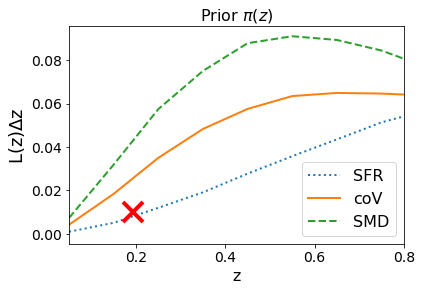}
    \includegraphics[width=0.45\textwidth]{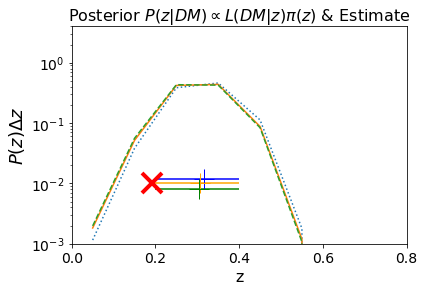}
    \caption{Example of the inference of host redshift for the localized Spitler-burst FRB121102, indicated by a red cross \citep{tendulkar2017host}.
        Top left: Expected likelihood $L(\DMEG|z)$ assuming FRBs from magnetars in our benchmark scenario (Sec. \ref{sec:models}) for increasing redshift, indicated by the colorbar, together with the extragalactic $\DM_{\rm Spitler} \approx 340 \unitDM$ inferred for the Spitler-burst.
        Top right: Values of $L(\DMEG|z)$ at $\DM_{\rm Spitler}$ for increasing $z$, renormalized to $1=\int L(z|\DM_{\rm Spitler}) \dd z$.
        Estimating the host redshift from this function implicitly assumes all redshifts to host FRBs with same probability.
        Bottom left: Prior $\pi(z)$ for host redshift (Sec. \ref{sec:FRBpoppy}) according to three assumed distributions and selection effects of Parkes (cf. Fig. \ref{fig:population}), that measured the displayed value of $\DM_{\rm Spitler}$.
        These are more realistic assumptions than uniform $\pi(z)$.
        Bottom right: Posterior $P(z|\DM_{\rm Spitler})$, Eq. \eqref{eq:bayes_DM}, for host redshift of the Spitler-burst for three assumed populations together with the expected host redshift and 1$\sigma$ standard deviation.
        The $z\approx 0.19$ of the localized Spitler-burst is on the edge of the 1$\sigma$ deviation.
        The high estimate on $z$ is probably due to an unlikely strong local contribution of DM, expected to accompany the observed $|RM| > 10^5 \unitRM$ signal.
        Mainly due to vast increase of the probed volume with redshift, the likelihood for the host to reside at $z < 0.1$ is lower by about a magnitude.
    }
    \label{fig:Spitler_estimate}
\end{figure*}

\subsubsection{Method}

Earlier work has estimated the redshift of FRBs, $\zFRB$, based on their DM \citep{dolag2015,Niino2018,luo2018,Pol2019}.
By comparing the likelihood $L(\DM|\zFRB)$ at different redshifts, upper limits on $\zFRB$ are obtained.
However, according to Bayes theorem (Eq.~\ref{eq:Bayes-theorem})
\begin{equation}
P(\zFRB|\DM) \propto L(\DM|\zFRB) ~\pi(\zFRB) ,
\label{eq:bayes_DM}
\end{equation}
these estimates can be improved by using the posterior $P(\zFRB|\DM)$ that considers a reasonable prior of source redshifts, $\pi(\zFRB)$.
Not accounting for this prior is equivalent to assuming the same number of FRBs from any redshift, thus ignoring distribution and evolution of FRBs, the telescopes selection effects as well as the fact, that the probed volume increases with distance.
The latter drastically lowers the amount of FRBs expected from low redshift $z \lesssim 0.2$, independent of the history of sources.
\citet{Walker2018} used a $\pi(\zFRB)$ deduced from the observed population of gamma-ray bursts and showed that this allows to obtain lower limits on $\zFRB$.
In Sec. \ref{sec:FRBpoppy}, we derive a better motivated
$\pi(\zFRB)$, considering intrinsic redshift distributions of FRBs as well as telescope selection effects.
By evaluating the contribution of each region along the LoS (see Secs. \ref{sec:IGM} - \ref{sec:local}), assuming FRBs from magnetars, we can estimate the distribution of extragalactic $\DMEG$.
We calculate the source redshift of FRBs by extracting the expectation value and 3$\sigma$-deviation from the posterior $P$ obtained by Eq. \eqref{eq:bayes_DM}.
In Fig. \ref{fig:Spitler_estimate} we show, as an example, the derivation of $\zFRB$ for the localized Spitler burst.
We obtain redshift estimates based on $\DMEG=\DMobs-\DMMW$ for all FRBs listed in the FRBcat \citep{FRBCAT}.
These values of $\DMEG$ were shown to be correct to $\approxeq 30 \unitDM$ \citep{manchester2005australia}. Results are shown in Table~\ref{tab:redshift_DM}.

\subsubsection{Results}

We estimate the redshift of the Spitler burst to be $z \approx 0.31$
Our over-estimate may be attributed to a strong local DM accompanying the high $\RM \gtrsim 10^5$ of FRB121102.
\\

We obtain 3$\sigma$ lower limits on the redshift of FRBs in FRBcat observed with $\DMEG \geq 400 \unitDM$, thus providing the first reasonable estimates on the host redshifts of a large set of unlocalized FRBs.
For comparison, \citet{Pol2019} derive lower limits for only a single FRB160102, observed with $\DM \approx 2596 \unitDM$.

\subsubsection{Discussion}

In order to derive the most conservative lower limits, we overestimate the intergalactic $\DMIGM$, by assuming all baryons to be localized in the ionized IGM, $\fIGM=1$, thus associating the same value of DM with lower redshifts than for smaller choices of $\fIGM$.
However, more realistic estimates should account for the conservation of baryons, which partly reside in collapsed regions along the LoS, thus $\fIGM \leq 0.9$ (cf. Sec. \ref{sec:IGM}).
\\

Since at low $z$ the redshift distribution of FRBs is dominated by the increase of the probed volume, rather than the history of the sources, the lower limits are consistent among the different assumed scenarios.
Lower values of $\DMEG \lesssim 400 \unitDM$ are more likely to be caused by the local environment or the host galaxy and can be explained by an FRB in the local Universe, thus do not allow for a lower limit on their redshift.
However, the local environment of magnetars in the local Universe have a very small chance ($\lesssim 0.02$  per cent in our model) to contribute $\DM > 10^3 \unitDM$, up to several $10^4 \unitDM$.
Thus $z=0$, can never be entirely excluded.
Still, the results obtained in this section can be used to estimate the distribution of FRB host redshifts from unlocalized events.

\begin{table*}
    \input{table_redshifts_DM.tex}
    \caption{Redshift estimates for 38 FRBs catalogued in FRBcat \citep{FRBCAT} with observed $\DMobs$ and estimated Galactic foreground $\DMMW$ with $\DMobs - \DMMW \gtrsim 400 \unitDM$ (exact number depends on observing telescope), for which we can estimate 3$\sigma$ lower limits (cf. Fig. \ref{fig:Spitler_estimate}).
        3$\sigma$ ranges are computed numerically and show the outer edges of the range that contains $>99.7$ per cent of probability, which yields conservative estimates, as an exact computation would result in a more narrow range.
        We obtain estimates assuming all baryons to be localized in the ionized IGM, $\fIGM = 1$, in order to arrive at the most conservative lower limits, since for lower $\fIGM$, the same value of $\DMEG$ is associated with further distance.
        We are able to obtain lower limits on the host redshift by applying Bayes theorem (Eq. \eqref{eq:Bayes}), combining the full likelihood $L(\DM|z)$, assuming FRBs from magnetars in our benchmark scenario (Sec. \ref{sec:models}), with a prior $\pi(z)$ on host redshift derived in Sec. \ref{sec:FRBpoppy}.
        Assuming different redshift distributions of FRBs, see Fig. \ref{fig:population}, does not affect the lower limits, since they all share the increase of the probed volume that dominates their shape at low redshift.
    }
    \label{tab:redshift_DM}
\end{table*}


\subsection{Inference of intergalactic magnetic field}
\label{sec:IGMF}
\begin{figure}
    \centering
    \includegraphics[width=0.5\textwidth]{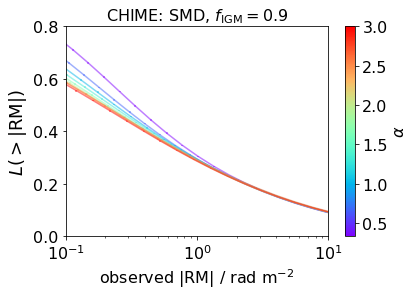}
    \includegraphics[width=0.5\textwidth]{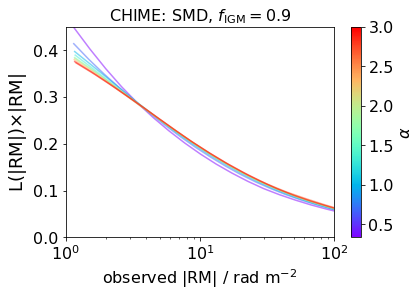}
    \caption{Complementary cumulative (top) and differential (bottom) distribution of $\RMEG$ expected to be observed by CHIME, assuming FRBs from magnetars in our benchmark scenario (Sec. \ref{sec:models}), their redshift distribution to follow SMD and an amount of baryons in the IGM, $\fIGM=0.9$.
    Colors indicate different choices for exponent $\alpha$ of the $B$-$\rho$-relation (Eq. \eqref{eq:B_rho}).
    The error bars that represent sampling shot-noise are barely visible, rendering the small difference significant.
    The amount of observable FRBs with $\RMEG \geq 10^{-1} \unitRM$ (top) as well as the renormalized distribution of reasonable $\RMEG > 1 \unitRM$ (bottom) is  influenced by the strength of IGMFs.
    This is true, independent of models chosen for the other regions along the LoS.
    $\RMEG > 10^2$ are almost completely determined by the local environment and thus not shown here.
    Our results show that the Spitler burst observed with $|\RM| > 10^5 \unitRM$ is a one-in-a-million source $L(> 10^5 \unitRM) \lesssim 10^{-6}$. However, due to its high rate of repetition, likelihood of detection is certainly much higher.
    }
    \label{fig:RM_telescope}
\end{figure}

\subsubsection{Method}

In this Section we discuss the use of the DM and RM of unlocalized FRBs to put constraints on the index $\alpha$ of $B$-$\rho$-relation in the IGM (cf. Eq. \ref{eq:B_rho} and Fig. \ref{fig:rho-B_relations}).
However, since the RM has the same dependency on the free electron density $n_e$ as DM, it is likewise affected by $\fIGM$ -- see Eqs. \eqref{eq:DM_IGM} - \eqref{eq:RM_IGM}.
We assume $\fIGM = 0.9$ in order to maximize the contribution of the IGM.

\subparagraph{Combined inference of DM and RM}

According to Eq.~\eqref{eq:combined}, the full information from, both, DM and RM of the same unlocalized event can be obtained as
\begin{equation}
L(\DM,~\RM|,~\alpha) 
=  \int \pi(z) ~ L(\DM |z) ~ L(\RM|z,\alpha)~ \dd z ,
\label{eq:combinedDMRM}
\end{equation}
thus delivering us the combined likelihood of $\fIGM$ and $\alpha$.
The likelihoods $L(\DM|z)$ and $L(\RM|z)$ represent our expectations for the extragalactic contribution to DM and RM, respectively, for FRBs produced at magnetars in our benchmark scenario that considers all regions along the LoS (see Sec. \ref{sec:models}), including intervening galaxies.
Eq. \eqref{eq:combinedDMRM} can be interpreted by identifying the RM-free part of the integrand with the posterior (Eq. \eqref{eq:bayes_DM}) shown in lower-right plot of Fig. \ref{fig:Spitler_estimate} , which quantifies our expectation for the host redshift based on DM of the individual unlocalized FRB.
This posterior, in turn, acts as the prior for host redshift when interpreting RM regarding the IGMF.
This detailed combined analysis of expected distribution of DM and RM for FRBs from different possible host redshift allows to obtain the full information entailed in the observables of FRBs.
By renormalizing $L(\DM,\RM)$ to the same choice of $\alpha$ for all events, we obtain the Bayes factor $\Bayes$ (Eq. \ref{eq:Bayes}).
Since we assume that all $\alpha$ have identical priors, $\pi(\alpha) = \rm const.$, $\Bayes$ is identical to the ratio of posteriors.

\subparagraph{Mock sample}

Here we estimate how many unlocalized CHIME\footnote{Note that we are mostly interested in $\RM \ll 10^3 \unitRM$, which can be probed at low frequencies \citep{fonseca2020}} FRBs are required in order to measure $\alpha$.
To this end, we produce mock samples of FRBs, sampling DM and RM according to estimates in our benchmark scenario (Fig. \ref{fig:RM_telescope}, Sec. \ref{sec:models}), assuming the weakest of IGMFs, i.e. $\alpha = 9/3$.
Investigation of the IGMF with unlocalized FRBs is degenerate to the host redshift distribution and $\fIGM$, preventing reasonable conclusions in a joint analysis.
We choose the SMD distribution which peaks at lowest redshift of the three compared distributions, thus provides the smallest IGM contribution to RM.
The required number of FRBs will hence be lower for the other distributions that peak at more distant redshift.
We further assume the maximum possible amount of baryons in the IGM, $\fIGM = 0.9$, as suggested by the Macquart relation \citep{Macquart_2020}.
By increasing the sample size $N_{\rm FRB}$, we investigate how many FRBs are required in order to rule out choices of $\alpha$, i.e. $\Bayes(\alpha) < 10^{-2}$.
For each value of $N_{\rm FRB}$, we take 10 samples, for which we compute the total value for $\Bayes$ and show the logarithmic mean and standard deviation in Fig.~\ref{fig:mock_alpha}.

\subsubsection{Results}

In Fig. \ref{fig:RM_telescope} we show the likelihood of $\RMEG$ to be observed by CHIME, assuming the redshift distribution of FRBs to follow SMD.
The top plot shows the likelihood of FRBs observed with $|\RMEG| > 0.1 \unitRM$ which decreases from 70.6 per cent for $
\alpha = \frac{1}{3}$ to 59.2 per cent for $\alpha=\frac{9}{3}$.
However, the number of observed FRBs expected to have $|\RMEG| > 1 \unitRM$ for $\alpha = \frac{1}{3}$ is 30.7 per cent and 29.5 per cent for $\alpha = \frac{9}{3}$, thus hard to distinguish.

Still, the lower $\alpha$, i.e. the stronger the IGMF, the more FRBs with $0.1 \unitRM \lesssim |\RMEG| < 10 \unitRM$ will be observed.
This qualitative result is independent on the exact model of IGMF or assumptions regarding the other regions.
Thus, the number of FRBs observed with significant $\RMEG$ in a survey with systematically extracted RM is a good indicator for the IGMF.
However, the expected likelihood of $|\RMEG| > 0.1 \unitRM$ will change when other models are considered and perhaps hamper the inference of the IGMF.
Note that the assumed models for local environment and host galaxy have a decent chance to provide $|\RMEG| < 0.1 \unitRM$, due to old magnetar ages or bimodal distribution of galactic magnetic fields with many virtually unmagnetized galaxies, thus allow for the inference of IGMFs.
The contribution of assumed models for the individual regions to the total observed signal can be seen in App. \ref{sec:telescope_predictions}.
This stresses how important it is to exactly estimate all contributions in order to correctly interpret the observed number and distribution of RM.
\\

The bottom plot of Fig. \ref{fig:RM_telescope} shows that the differential change in the amount of $\RMEG$ significantly changes the distribution of $|\RMEG| > \RM_{\rm min} = 1 \unitRM$, which can be used to infer $\alpha$ from this sub-sample only.
Hence, data with carefully subtracted galactic foregrounds can be used to constrain the IGMF.
Note that we assume $|\RMEG| > \RM_{\rm min} = 1 \unitRM$ can be inferred with precision of $1 \unitRM$, determined by the minimal range of bins, by removing the Galactic foreground, e.g. using a Wiener filter \cite{oppermann2015,hutschenreuter2020}.

However, the results in Fig. \ref{fig:RM_telescope} show differences beyond the statistical noise even if we choose higher minimum accessible values of $\RMEG$, $1 \unitRM \lesssim \RM_{\rm min} < 10 \unitRM$.
Thus, constraints on $\alpha$ might also be possible if the MW foreground can be removed with slightly worse precision than $1 \unitRM$.
This stresses the importance of reliable estimates of the Galactic contribution to the RM as well as confirming the results of Galactic foreground filters with robust models for the density and magnetic field of the MW \citep{imagine_whitepaper}.
\\

\begin{figure}
    \centering
    \includegraphics[width=0.5\textwidth]{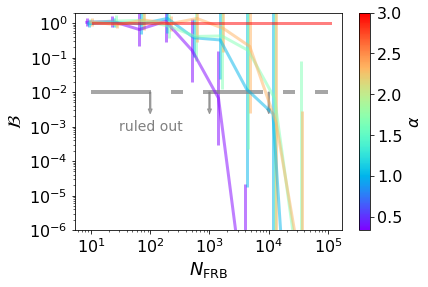}
    \caption{Bayes factor $\Bayes$ for different values of $\alpha$ for mock samples of FRBs with increasing size $N_{\rm FRB}$ assumed to be observed by CHIME in our benchmark scenario assuming FRBs from magnetars, the weakest IGMF model ($\alpha=\frac{9}{3}$), a redshift distribution following SMD, as well as $\fIGM = 0.9$.
            The error bars show the standard deviation for the results of 10 samples of similar size.
            $\Bayes$ factors for all $\alpha$ compare to the case of $\alpha=\frac{9}{3}$, thus $\Bayes(\alpha_0) < 1e-2$, marked by the gray line, are considered decisive to rule out $\alpha_0$.
            The transition of $\Bayes(N_{\rm FRB}|\alpha_0)$ through that line marks the minimum required number of FRBs observed with $\RMEG> 1 \unitRM$ to constrain $\alpha > \alpha_0$.
                }
    \label{fig:mock_alpha}
\end{figure}

Fig. \ref{fig:mock_alpha} shows that at least $N_{\rm FRB}=10^3$ FRBs observed with $\RMEG \geq 1 \unitRM$, which is $\lesssim 1/3$ of all events, are required in order to constrain $\alpha < \frac{1}{3}$, i.e. constraints comparable to the current upper limit \citep[$B < 4.4 ~\rm nG$][]{Planck2015PMF}.
Moreover, for $N_{\rm FRB} \gtrsim 5 \times 10^4$, most $\alpha \leq 8/3$ are ruled out, allowing to probe the IGMF down to the current lower limit \citep[$B > 3\times10^{-7} ~\rm nG$,][]{neronov2010}.

However, in order to infer the IGMF down to the limit by \citet{neronov2010}, a much greater sample is required than these telescopes can acquire in a life-time.
Instead, this requires large arrays of telescopes that systematically observe several thousand FRBs each year -- such as the SKA \citep{macquart2015fast}.
Furthermore, the presented estimates on $N_{\rm FRB}$ are optimistic and depend on the exact modelling of all regions along the LoS, which need to be verified by other observables.

\subsubsection{Discussion}

By using the high value of $\fIGM = 0.9$, we obtain the most optimistic estimate for $N_{\rm FRB}$.
For lower values of $\fIGM$, $\RMIGM$ is reduced and a lower number of LoS will be able to significantly contribute to detectable RM.  This in turn might increase the number of FRBs $N_{\rm FRB}$, necessary to constrain $\alpha$,  and this will also decrease the range of $\alpha$ detectable using the RM.

Moreover, the ensemble used to model the host and the intervening galaxies contains a significant number of galaxies that do not meet conditions for large-scale dynamos,  and thus can only carry weak coherent magnetic fields (cf. Sec. \ref{sec:host}).
This results in a rather low RM  contribution from these regions, compared to other works \citep[e.g.][]{Basu2018}.
The galaxy models are considered to a distance, at which the surface mass density falls to 1 per cent of the central value, and thus do not account for the halo of galaxies, 
However, the sources of FRBs might be located at the edge of their host galaxies, if there is sufficient molecular gas to indicate star formation.
Such short LoS, especially within the numerous low-mass $M_\star \gtrsim 10^7 M_\odot$ galaxies, only contribute little to the DM and RM.
However, we implicitly assume that most FRBs reside in MW-like galaxies, which contain most stellar mass.
Still, by considering the numerous low-mass central galaxies of any possible brightness in the low density Universe, the model accounts for even weaker, though arguably more realistic estimates of the galaxy contributions as compared to other works.

Moreover, the elliptical galaxies in the \Rodrigues~ sample only account for negligible contributions to RM as only the vanishing large-scale magnetic field is considered for computation.
However, \citet{moss1996turbulence} suggest that high values of RM, up to $100 \unitRM$, might possibly be observed from ellipticals with sufficient resolution, which prevents the beam width to contain many correlation lengths whose Faraday rotation interfere destructively.
The small angular extent of FRBs renders their RM independent of the instruments angular resolution and hence might carry even higher values of RM from their elliptical host.
Future works should thus consider a more realistic estimate of the contribution from turbulent magnetic field in elliptical galaxies.
Overall, the low strength of coherent magnetic fields predicted by \citet{rodrigues2018} implies  that our conclusions on the IGMF are optimistic (see Sec. \ref{sec:IGMF}).

Furthermore, the contribution of the local environment is not well constrained and can significantly affect the shape of $L(\RMEG)$, which might be misinterpreted as signal of the IGMF.
In App. \ref{sec:telescope_predictions} we provide a comparison of the contributions of different regions to the observed distribution of measures.
This shows that basically all regions along the LoS provide significant amounts of RM.
Though we could show that RM of FRBs carry detailed information on IGMFs, we might not be able to extract this information, owing to the imprecise knowledge of foregrounds.
This stresses the importance to investigate FRBs with identified hosts, whose contribution can be estimated more precisely, as well as to identify the source of FRBs to more exactly quantify the contribution of the local environment.
However, even under these circumstances, the contributions of regions different than the IGM may hardly be known with required precision.
In future works we will consider further models for the other regions along the LoS in order to identify model-independent signals of the IGMF .

Unambiguous identification of IGMFs solely via $L(\RMEG)$ of FRBs requires realistic modelling of all contributions and an exact fit to the observed distribution.
However, there might be several fitting scenarios that consider different models.
Distinguishing between those solutions requires their verification using other measures of FRBs or different astrophysical signals.
In future works we aim to include more measures in \PreFRBLE, especially propagation-independent measures that carry information about the source.
\\

Note that the results in this section exclude Galactic contributions to the RM.
In order to constrain IGMFs, we need to be sensitive for $\RMEG \lesssim \text{few} \unitRM$. Hence, future work should account for RM foregrounds due to the MW as well as the ionosphere.

However, the estimate of $N_{\rm FRB}$ are not affected by the Galactic foregrounds that we assume can be removed with precision of $1 \unitRM$ to identify extragalactic components $\RMEG \geq 1 \unitRM$.

%% file: table_redshifts_DM.tex
\begin{tabular}{l|c|c|c|c|c|c}
	ID & $\DMobs$ / $\unitDM$  & $\DMMW$ / $\unitDM$ & $z_{\rm SFR}(\DM)$  & $z_{\rm coV}(\DM)$  & $z_{\rm SMD}(\DM)$ \\
	FRB190604 & 552.7 & 32.0 & $0.54 _{-0.44} ^{+0.36}$ & $0.52 _{-0.42} ^{+0.38}$ & $0.51 _{-0.41} ^{+0.39}$ \\
	FRB190417 & 1378.1 & 78.0 & $1.28 _{-0.78} ^{+0.72}$ & $1.24 _{-0.84} ^{+0.76}$ & $1.19 _{-0.79} ^{+0.81}$ \\
	FRB190222 & 460.6 & 87.0 & $0.39 _{-0.29} ^{+0.31}$ & $0.37 _{-0.27} ^{+0.33}$ & $0.37 _{-0.27} ^{+0.33}$ \\
	FRB190212 & 651.1 & 43.0 & $0.62 _{-0.42} ^{+0.38}$ & $0.60 _{-0.40} ^{+0.40}$ & $0.59 _{-0.49} ^{+0.41}$ \\
	FRB190209 & 424.6 & 46.0 & $0.39 _{-0.29} ^{+0.31}$ & $0.37 _{-0.27} ^{+0.33}$ & $0.37 _{-0.27} ^{+0.33}$ \\
	FRB190208 & 579.9 & 72.0 & $0.52 _{-0.42} ^{+0.38}$ & $0.50 _{-0.40} ^{+0.40}$ & $0.50 _{-0.40} ^{+0.30}$ \\
	FRB190117 & 393.3 & 48.0 & $0.36 _{-0.26} ^{+0.24}$ & $0.34 _{-0.24} ^{+0.26}$ & $0.34 _{-0.24} ^{+0.26}$ \\
	FRB190116 & 444.0 & 20.0 & $0.44 _{-0.34} ^{+0.26}$ & $0.42 _{-0.32} ^{+0.28}$ & $0.42 _{-0.32} ^{+0.28}$ \\
	FRB181017 & 1281.9 & 43.0 & $1.22 _{-0.72} ^{+0.68}$ & $1.18 _{-0.78} ^{+0.72}$ & $1.14 _{-0.74} ^{+0.76}$ \\
	FRB180817 & 1006.8 & 28.0 & $0.98 _{-0.58} ^{+0.62}$ & $0.94 _{-0.64} ^{+0.66}$ & $0.92 _{-0.62} ^{+0.58}$ \\
	FRB180812 & 802.6 & 83.0 & $0.73 _{-0.43} ^{+0.47}$ & $0.70 _{-0.50} ^{+0.50}$ & $0.69 _{-0.49} ^{+0.41}$ \\
	FRB180806 & 740.0 & 41.0 & $0.71 _{-0.51} ^{+0.49}$ & $0.68 _{-0.48} ^{+0.42}$ & $0.67 _{-0.47} ^{+0.43}$ \\
	FRB180801 & 656.2 & 90.0 & $0.58 _{-0.38} ^{+0.42}$ & $0.56 _{-0.46} ^{+0.34}$ & $0.55 _{-0.45} ^{+0.35}$ \\
	FRB180730 & 849.0 & 57.0 & $0.80 _{-0.50} ^{+0.50}$ & $0.77 _{-0.57} ^{+0.53}$ & $0.76 _{-0.56} ^{+0.54}$ \\
	FRB180727 & 642.1 & 21.0 & $0.63 _{-0.43} ^{+0.37}$ & $0.61 _{-0.41} ^{+0.39}$ & $0.60 _{-0.50} ^{+0.40}$ \\
	FRB180725 & 716.0 & 71.0 & $0.66 _{-0.46} ^{+0.44}$ & $0.63 _{-0.43} ^{+0.47}$ & $0.62 _{-0.42} ^{+0.48}$ \\
	FRB180714 & 1467.9 & 257.0 & $1.21 _{-0.71} ^{+0.69}$ & $1.17 _{-0.77} ^{+0.73}$ & $1.13 _{-0.73} ^{+0.77}$ \\
	FRB180311 & 1570.9 & 45.2 & $1.50 _{-0.90} ^{+0.90}$ & $1.47 _{-0.87} ^{+0.93}$ & $1.41 _{-0.91} ^{+0.89}$ \\
	FRB171209 & 1457.4 & 13.0 & $1.43 _{-0.83} ^{+0.87}$ & $1.39 _{-0.89} ^{+0.91}$ & $1.34 _{-0.84} ^{+0.86}$ \\
	FRB160102 & 2596.1 & 13.0 & $2.45 _{-1.25} ^{+1.65}$ & $2.53 _{-1.43} ^{+1.77}$ & $2.31 _{-1.51} ^{+1.69}$ \\
	FRB151230 & 960.4 & 38.0 & $0.93 _{-0.53} ^{+0.57}$ & $0.90 _{-0.60} ^{+0.60}$ & $0.88 _{-0.58} ^{+0.62}$ \\
	FRB151206 & 1909.8 & 160.0 & $1.70 _{-0.90} ^{+1.00}$ & $1.68 _{-1.08} ^{+1.12}$ & $1.59 _{-0.99} ^{+1.11}$ \\
	FRB150610 & 1593.9 & 122.0 & $1.45 _{-0.85} ^{+0.85}$ & $1.42 _{-0.92} ^{+0.88}$ & $1.36 _{-0.86} ^{+0.84}$ \\
	FRB150418 & 776.2 & 188.5 & $0.60 _{-0.40} ^{+0.40}$ & $0.58 _{-0.48} ^{+0.42}$ & $0.57 _{-0.47} ^{+0.43}$ \\
	FRB150215 & 1105.6 & 427.2 & $0.69 _{-0.49} ^{+0.41}$ & $0.66 _{-0.46} ^{+0.44}$ & $0.65 _{-0.45} ^{+0.45}$ \\
	FRB140514 & 562.7 & 34.9 & $0.54 _{-0.44} ^{+0.36}$ & $0.52 _{-0.42} ^{+0.38}$ & $0.51 _{-0.41} ^{+0.39}$ \\
	FRB131104 & 779.0 & 71.1 & $0.72 _{-0.52} ^{+0.48}$ & $0.69 _{-0.49} ^{+0.41}$ & $0.68 _{-0.48} ^{+0.42}$ \\
	FRB130729 & 861.0 & 31.0 & $0.85 _{-0.55} ^{+0.45}$ & $0.81 _{-0.61} ^{+0.49}$ & $0.80 _{-0.60} ^{+0.50}$ \\
	FRB130628 & 469.9 & 52.6 & $0.42 _{-0.32} ^{+0.28}$ & $0.41 _{-0.31} ^{+0.29}$ & $0.41 _{-0.31} ^{+0.29}$ \\
	FRB130626 & 952.4 & 66.9 & $0.90 _{-0.60} ^{+0.50}$ & $0.86 _{-0.56} ^{+0.54}$ & $0.84 _{-0.54} ^{+0.56}$ \\
	FRB121002 & 1629.2 & 74.3 & $1.53 _{-0.83} ^{+0.87}$ & $1.50 _{-0.90} ^{+1.00}$ & $1.44 _{-0.94} ^{+0.96}$ \\
	FRB120127 & 553.3 & 31.8 & $0.54 _{-0.44} ^{+0.36}$ & $0.52 _{-0.42} ^{+0.38}$ & $0.51 _{-0.41} ^{+0.39}$ \\
	FRB110703 & 1103.6 & 32.3 & $1.08 _{-0.68} ^{+0.62}$ & $1.04 _{-0.74} ^{+0.66}$ & $1.01 _{-0.71} ^{+0.69}$ \\
	FRB110626 & 723.0 & 47.5 & $0.69 _{-0.49} ^{+0.41}$ & $0.66 _{-0.46} ^{+0.44}$ & $0.65 _{-0.45} ^{+0.45}$ \\
	FRB110220 & 944.4 & 34.8 & $0.93 _{-0.53} ^{+0.57}$ & $0.89 _{-0.59} ^{+0.61}$ & $0.87 _{-0.57} ^{+0.53}$ \\
	FRB090625 & 899.5 & 31.7 & $0.88 _{-0.58} ^{+0.52}$ & $0.84 _{-0.54} ^{+0.56}$ & $0.83 _{-0.63} ^{+0.57}$ \\
	FRB010312 & 1187.0 & 51.0 & $1.14 _{-0.64} ^{+0.66}$ & $1.10 _{-0.70} ^{+0.70}$ & $1.07 _{-0.77} ^{+0.73}$ \\
	FRB010125 & 790.0 & 110.0 & $0.70 _{-0.50} ^{+0.40}$ & $0.67 _{-0.47} ^{+0.43}$ & $0.66 _{-0.46} ^{+0.44}$ \\
\end{tabular}

%% file: discussion.tex

\section{Conclusion}
\label{sec:discussion}

We model the extragalactic contribution to propagation dependent measures of FRBs from all regions along the line-of-sight, including intervening galaxies, as well as the distribution of host redshift.
Using approximate Bayesian computation, we estimate the expected distribution of disperion measure, rotation measure and temporal smearing $\tau$ and compare to observations.
Our code is provided as an open-source Python software package \PreFRBLE\footnote{\href{https://github.com/FRBs/PreFRBLE}{github.com/FRBs/PreFRBLE}} \citep{PreFRBLEzenodo}.
\\

We use \PreFRBLE ~to identify intervening galaxies, estimate the host redshift of unlocalized FRBs and infer the strength of the intergalactic magnetic field.

The main conclusions of our work can be summarized as follows:

\begin{itemize}
    \item We find that intervening galaxies are unlikely to account for the entirety of high values of temporal smearing $\tau > 0.06 ~\rm ms$ observed by the Parkes Telescope.
    The most likely explanation is the presence of a denser and more turbulent environment of the magnetar progenitor than expected from Galactic magnetars. 
    This is in line with earlier  findings by \citet{margalit2020implications}.
    \item By applying Bayes theorem and making physically motivated assumptions on the redshift distribution of FRBs, we obtain realistic $3\sigma$ lower limits on the estimated redshift, from a big sample of 38 unlocalized FRBs with $\DMEG \gtrsim 400 \unitDM$ (see Tab.~\ref{tab:redshift_DM}), independent on the exact shape of assumed distribution of host redshift, and thus also independent of source model and history.
    \item The stronger the IGMF, the more FRBs with $0.1 \unitRM \lesssim |\RMEG| < 10 \unitRM$ will be observed.
    This is independent of assumptions regarding the other regions and on the exact model of IGMF.
    However, the exact number of FRBs with such $\RMEG$ is also influenced by the details of other regions along the LoS, which all contribute significant amounts of RM and maybe hamper the inference of IGMFs.
    In order to arrive at reasonable conclusions, more competing models have to be considered for each of these regions.
    
    \item In order to put constraints on the strength of IGMFs of the order of the current upper limit \citep[$B\lesssim 4.4 \times 10^{-9} \rm ~G$,][]{Planck2015PMF}, we predict that a number of at least $10^3$ unlocalized FRBs with associated $\RMEG > 1\unitRM$ (i.e. $\lesssim 1/3$ of all events) is required to be observed with CHIME. For this estimate we assumed  $\fIGM = 0.9$, and that Galactic foregrounds can be removed with a precision of $\leq 1 \unitRM$.
   Furthermore, it will be possible to derive constraints of the order of the current lower limits from entirely different proxies \citep[$B \gtrsim 3\times 10^{-16} ~\rm G$,][]{neronov2010}, once a higher number of events,  $\gtrsim 5 \times 10^4$ FRBs, have been observed with $\RMEG > 1\unitRM$.
\end{itemize}

Our estimates of the extragalactic DM, RM and $\tau$ make use of models for the local environment of the source, the ensemble of host galaxies, the IGM as well as the ensemble of intervening galaxies.
However, we ignore the clustering or haloes of host and intervening galaxies \citep[e. g.][]{prochaska2019low,connor2020} and do not account for foregrounds from the MW, the Galactic Halo or Earth's ionosphere, which we assume can be removed with sufficient precision to infer the extragalactic component. This will be the subject of future work.

%% file: appendix.tex



\appendix

\section{Compare inner scale to plasma phase fluctuations }
\label{sec:rdiff}
\begin{figure}
\centering
\includegraphics[width=0.5\textwidth]{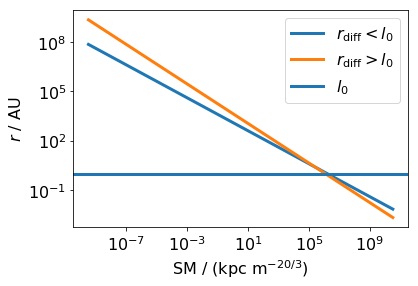}
\caption{$\rdiff$ as function of contributed $\SMeff$ for wavelength $\lambda_0=0.23 \rm~m$, prodived by \citet{macquart2013temporal}.
    This is used to distinguish between cases of phase structures with different solutions for $\tau$.
The blue function shows results valid below $l_0$, orange is valid above $l_0$.
$l_0 = 1\rm~ AU $ is marked by the horizontal line.
For the models used in this work, all values of $\SMeff< 10^5 \unitSM$.
We thus only need to consider results for the case, where $l_0 > r_{\rm diff}$.
}
\label{fig:rdiff}
\end{figure}
\citet{macquart2013temporal} provide numerical expressions for the temporal scatter $\tau$ for plasma phase fluctuations $\rdiff$ larger or shorter than the inner scale of turbulence $l_0$.
In order to distinguish between these cases, we use their Eq. 10 to compute $\rdiff$ and compare to $l_0=1 \rm ~AU$, as commonly assumed for the ISM.
Results are shown in Fig. \ref{fig:rdiff}.
Clearly, for all values of $\SMeff$ we find that $\rdiff > l_0$ is a very good assumption.
This way, the wavelength of FRB signal $\lambda_0$ is a global factor to the different contributions and other choices of $\lambda_0$ can easily be investigated in post-processing.


\section{Telescope predictions}
\label{sec:telescope_predictions}
\begin{figure*}
    \centering
    \includegraphics[width=\textwidth]{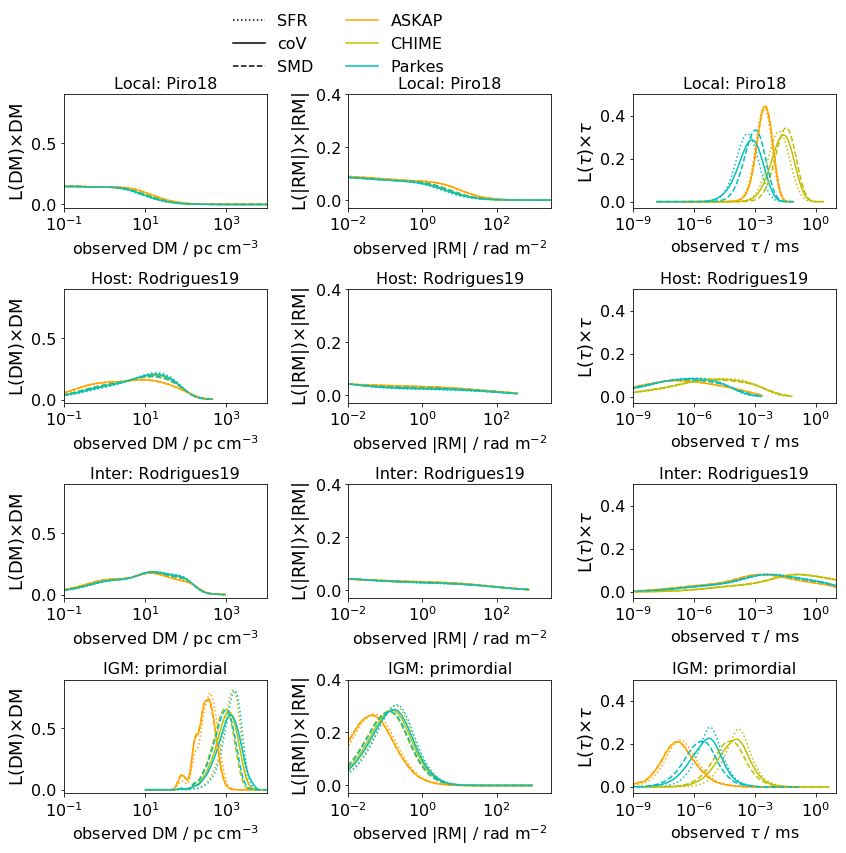}
    \caption{Likelihood of contribution to DM (left), RM (center) and $\tau$ (right) to FRBs by individual regions along LoS, indicated in the title, considering models in our benchmark scenario (Secs. \ref{sec:IGM} - \ref{sec:local}).
        Assuming the redshift distribution of FRBs to follow coV (solid), SMD (dashed) or SFR (dotted), distributions show the likely contribution of the model according to Eq. \eqref{eq:tele} in surveys conducted by individual telescopes, i. e. CHIME (blue), ASKAP (orange) and Parkes (green, see Sec. \ref{sec:FRBpoppy}).
        The $x$-axis is cut for values not accessible by terrestrial telescopes, while the shown distributions are normalized to 1 over the whole range of $x$.
        Comparing these plots allows to easily identify the dominant / significant / negligible regions and their contribution to observed values.
    }
    \label{fig:telescope_contributors}
\end{figure*}

Considering an intrinsic distribution of FRB host redshifts and selection effects (Sec. \ref{sec:FRBpoppy}) allows to predict the distribution of measures expected for individual telescopes.
In Fig. \ref{fig:telescope_contributors} we show the contribution of separate regions to the observed $\DM$, $\RM$ and $\tau$ for different telescopes and intrinsic redshift distributions.
These are obtained from Eq. \eqref{eq:tele}, using for $L(v|z)$ the predictions of individual models explained in Secs. \ref{sec:IGM} - \ref{sec:local}.
These plots allow to easily identify the regions that dominate a given measure.

\section{Effectively contributing over-density environments}
\label{sec:RM_overdensity}
\begin{figure*}
\centering
\includegraphics[width=0.9\textwidth]{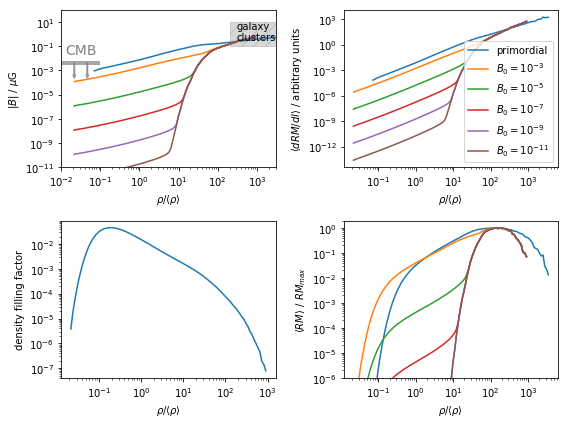}
\caption{Upper left: Median magnetic field strength $|B|$ as function of over-density $\rho/\avg{\rho}$ for a number of MHD models with identical dynamo physics, starting with different strengths of the primordial magnetic field $B_0$, indicated by the label in $\rm\mu G$. 
Upper right: average contribution to $\avg{\RM}$ per unit length as function of over-density. Obtained by multiplying $|B|(\rho)\times\rho/\avg{\rho}$ in the upper-left panel.
Lower left: model independent density volume filling factor $f(\rho)$.
Lower right: average contribution of $\avg{\RM}$ from regions in the IGM with different over-density, which is obtained by multiplying $\avg{\dd\RM} \times f(\rho)$ (upper right times lower left). 
Some of the models are not visible in that plot as they are identical to the case of $B_0 = 10^{-11} \rm~\mu G$ and the y-axis is set to only show the relevant contribution to $\avg{\RM}$. 
Models that are identical in this plot cannot be distinguished by investigation of $\avg{\RM}$.
}
\label{fig:RM_overdensity}
\end{figure*}
We want to use the RM data to distinguish between different models for the origin of IGMFs.
Ideally, one would hope to obtain constraints on the strength of the primordial magnetic field $B_0$, produced at $z\gg10$, by measuring the magnetic fields in voids.
Hence, we produced a number of MHD simulations that consider identical dynamo physics, but start with different strengths of the primordial magnetic field.
In Fig.~\ref{fig:RM_overdensity}, upper left, we show the $|B|\sim \rho$ relation obtained for the present day.
To compare the average contribution to $\avg{\RM}$ from regions with different densities, we estimate the contribution per unit length by $\avg{\frac{\dd RM}{\dd l}} \propto \rho |B|$, shown in the upper right plot.
We multiply this result by the average relative path length, approximated by the model independent density volume filling factor, shown in the lower left, in order to approximate the average contribution to $\avg{\RM}$ from these regions, shown in the lower right of Fig.~\ref{fig:RM_overdensity}. 

This plot clearly shows that even for the strongest primordial magnetic fields allowed by present constraints \citep{Planck2015PMF}, contributions to $\avg{\RM}$ are negligible from regions with over-densities below 0.1, where the comoving primordial magnetic field strength might be conserved. 
Hence, $\avg{\RM}$ is not a direct probe of the magnetic field in voids or of the primordial magnetic field.

However, for over-densities associated with filaments and sheets, $10 < \rho/\avg{\rho} < 200$, different shapes of $|B|(\rho)$ can have a significant impact on $\avg{\RM}$.
Moreover, the statistical distribution of RM, expressed by the likelihood function $L(\RM)$, allows for a more precise investigation of RM along different LoS, especially of paths that do not enter high-density regions, which dominate $\avg{\RM}$.
This way, $L(\RM)$ can be used to constrain parts of $|B|(\rho)$ to lower values of $\rho/\avg{\rho} < 10$ than by using $\avg{\RM}$, allowing to obtain general conclusions on models for the IGMF, independent of the individual formation processes.


\section{Path length of LoS through ellipsoid}
\label{sec:HostPathLength}
\begin{figure}
\centering
\includegraphics[width=0.5\textwidth]{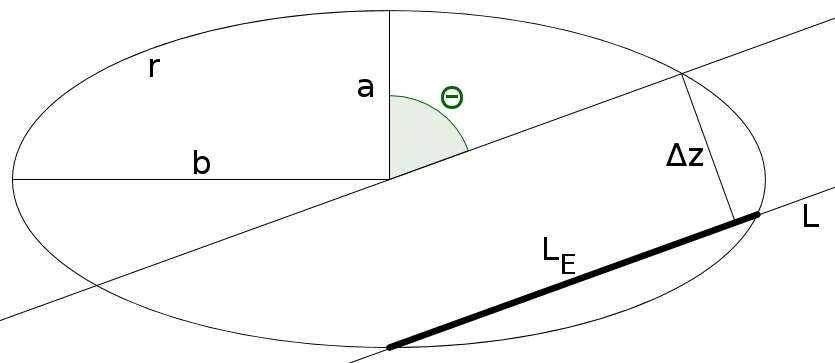}
\caption{Path length $L_E$ of LoS $L$ through ellipse $r$ viewed at angle $\theta$ with offset $\Delta z$ from the center.}
\label{fig:ellipse}
\end{figure}

We want to calculate the path length of a LoS within an axisymmetric galaxy model, represented by a three-dimensional ellipsoid with two identical axes $b_G$, thus
\begin{equation}
1 = \frac{x^2}{a_G^2} + \frac{y^2}{b_G^2} + \frac{z^2}{b_G^2},
\label{eq:ellipsoid}
\end{equation}
where x coordinate points along the major axis $a_G$ while y and z are rotationally invariant.
Viewed face on with inclination angle $\theta = 0$, impact parameters $\Delta y$ and $\Delta z$ both correspond to offset of the LoS from the center of the disc along axes y and z, respectively.
However, for $\theta \neq 0$, $\Delta z$ also entails an offset from the mid-plane at the center of the disc, while only $\Delta y$ still corresponds to the y-axis of the ellipsoid.
Since the y-coordinate is completely determined by $\Delta y$, it suffices to consider a LoS passing an ellipse.
Eq. \eqref{eq:ellipsoid} can be rewritten, such that
\begin{equation}
1 = \frac{x^2}{a_G^2 \left( 1 - \frac{y^2}{b_G^2} \right) }  + \frac{z^2}{b_G^2\left( 1 - \frac{y^2}{b_G^2} \right)} ,
\label{eq:ellipse}
\end{equation}
which defines an ellipse with axes
\begin{eqnarray}
a & = & a_G \sqrt{\left( 1 - \frac{y^2}{b_G^2} \right)} \\
b & = & b_G \sqrt{\left( 1 - \frac{y^2}{b_G^2} \right)}.
\label{eq:axes}
\end{eqnarray}
To obtain the path length $L_E$ of LoS $L$ through ellipse $r$ viewed at angle $\theta$ with offset $\Delta z$ from the center (see Fig. \ref{fig:ellipse}), we parametrize
\begin{equation}
L(t|\theta) = t \begin{pmatrix} \sin \theta \\ \cos \theta \end{pmatrix} + \Delta \begin{pmatrix} \cos \theta \\ - \sin \theta \end{pmatrix}
\end{equation}
and solve for intersections with ellipse
\begin{equation}
r(\tau) = \begin{pmatrix} b \sin\tau \\ a \cos\tau \end{pmatrix} ,
\end{equation}
 found at
\begin{equation}
\tau_{1/2} = - 2 \operatorname{atan}{\left(\frac{b \cos{\left(\theta \right)} \pm \sqrt{a^{2} \sin^{2}{\left(\theta \right)} + b^{2} \cos^{2}{\left(\theta \right)} - z^{2}}}{a \sin{\left(\theta \right)} - z} \right)} .
\end{equation}
This delivers the path length of the LoS within the ellipsoid
\begin{equation}
L_E = \sqrt{a^{2} \left(\cos{\left(\tau_{1} \right)} - \cos{\left(\tau_{2} \right)}\right)^{2} + b^{2} \left(\sin{\left(\tau_{1} \right)} - \sin{\left(\tau_{2} \right)}\right)^{2}} .
\end{equation}